\documentclass[twocolumn,secnumarabic,amssymb, nobibnotes, aps, pra]{revtex4-2}

\usepackage{amsmath,amssymb,amsfonts,color,epsfig}
\usepackage{bm}
\usepackage{graphicx}
\usepackage{subfigure}
\usepackage{setspace}
\usepackage{mathtools}
\usepackage{bm}
\usepackage{physics}
\usepackage{yfonts}
\usepackage{tikz-cd}
\usepackage{tikz}
\usepackage{booktabs}
\usepackage{tabularx}
\usetikzlibrary{decorations.markings}
\usepackage{framed}
\usetikzlibrary{shapes.geometric, arrows, positioning}
\usepackage[T1]{fontenc}
\usepackage{hyperref}
\hypersetup{
    colorlinks=true,
    linkcolor=blue,
    citecolor=blue,
    urlcolor=blue
}

\newcommand{\nn}{\nonumber}

\begin{document}
\title{Quantum Fisher information and quadrature squeezing in Janus superpositions of squeezed vacua}

\author{Arash Azizi}
    \affiliation{Texas A\&M University, College Station, TX 77843}

\begin{abstract}
Janus states, defined as coherent superpositions of two single-mode squeezed vacua, provide a simple but genuinely non-Gaussian setting for studying how interference reshapes quantum Fisher information (QFI) beyond the Gaussian squeezed-vacuum picture. Using an exact analytic treatment, we determine the QFI of Janus states and identify the benchmarks under which they can or cannot offer a metrological advantage over the single squeezed vacuum. We find that, under a fair comparison at fixed mean photon number, the single squeezed vacuum remains optimal for principal second-moment squeezing, so no genuine Janus advantage exists at that level. By contrast, within a fixed two-state span, a Janus superposition can simultaneously outperform its constituents in a laboratory quadrature variance and in number-generated phase QFI. We also introduce an operational benchmark based on fixed measured squeezing and show that, at the same observed squeezing level, Janus interference can substantially enhance the QFI for quadratic-generator sensing beyond the pure-Gaussian squeezed-vacuum reference. These results show that the metrological performance of Janus states is controlled not only by quadrature squeezing, but also by higher-order fluctuations and by the benchmark used for comparison.
\end{abstract}

\maketitle
\thispagestyle{empty}

\section{Introduction}
\label{sec:intro}

Precision sensing lies at the heart of modern quantum science. In a typical metrological task, one seeks to estimate a small physical parameter---such as a phase, frequency, coupling strength, displacement, or loss coefficient---from measurements on a probe whose quantum state depends on that parameter. Quantum metrology provides the framework for this problem by identifying the ultimate precision allowed by quantum mechanics and by clarifying when nonclassical probes can outperform classical ones under a specified resource constraint \cite{Braunstein_Caves1994,Giovannetti2004quantum_enhanced_measurements,Giovannetti2006,Paris2009quantum_estimation,Pezzesmerzi2018RMP,Toth_Apellaniz2014,DemkowiczDobrzanski2015optical_interferometry,Braun2018metrology_without_entanglement}. In the pure-state unitary setting, the central figure of merit is the quantum Fisher information (QFI), which is equal to four times the variance of the generator that encodes the parameter. The QFI therefore serves both as a rigorous precision benchmark and as a practical language for comparing probe families, sensing generators, and resource costs.

This perspective is especially natural in quantum optics, where the probe is often a bosonic mode and the relevant fluctuations can be shaped through squeezing, interference, and non-Gaussian superposition \cite{Mandel_Wolf1995,Scully_Zubairy1997,Loudon_Knight1987squeezed,RevModPhys.92.035005}. The canonical Gaussian resource is the single-mode squeezed vacuum, whose reduced quadrature noise has become a standard tool in precision measurement \cite{caves1981quantum,walls1983squeezed,Yurke1986,Hudelist2014,Sahota2016phase_estimation}. Experimentally, squeezed light has advanced from early optical demonstrations to high-purity modern platforms and large-scale interferometric implementations \cite{slusher1985observatio,Wu1986generation,Breitenbach1997,vahlbruch2008observation,mehmet2011squeezed,vahlbruch2016detection,aasi2013enhanced,Oelker:14,LIGO2016,Schnabel2017laser_interferometers}. At the same time, these developments have made an important point increasingly clear: reduced quadrature noise, nonclassical photon correlations, and enhanced QFI are closely related, but they are not identical notions of metrological advantage.

Superpositions of squeezed states have long been part of the quantum-optical landscape \cite{Sanders1989Superposition, Buzek1992Superpositions, Obada1997,
Obada_Al-Kader1999, ElOrany1999}. The Janus program revisited this class from a more specific perspective, namely the role of exact optical coherence in non-Gaussian interference. The original Janus work derived an analytic expression for \(g^{(2)}\) for a coherent superposition of two squeezed vacua and showed that interference can strongly suppress two-photon events while still obeying the universal lower bound \(g^{(2)} \ge 1/2\), with a practical minimum reached at moderate squeezing \cite{Azizi2026Janus}. A later study extended this analysis to arbitrary-order coherence by deriving exact expressions for \(g^{(k)}\), thereby showing that Janus interference can act as a tunable switch for multiphoton correlations through a squeezing-polynomial framework \cite{Azizi2025Janus_higher}. The displaced Janus generalization then broadened the construction to superpositions of squeezed coherent states with a common displacement and developed a wider analytical toolbox that included factorial moments, Wigner functions, non-Gaussianity diagnostics, and QFI for linear and quadratic generators \cite{Azizi2025displacedJanus}. More recently, the same program was extended to the two-mode setting through the Two-Mode Janus State (TMJS), defined as a coherent superposition of two distinct two-mode squeezed states and developed as a direct non-Gaussian generalization of the thermofield double, with exact phase-steerable higher-order coherence and a corresponding two-mode squeezing-polynomial structure \cite{Azizi2025TMJS}.

Those results established the coherence-theoretic and non-Gaussian foundations of the Janus family, but they did not settle the metrological question studied here. Once the discussion moves from \(g^{(2)}(0)\) and \(g^{(k)}(0)\) to QFI, different issues become central. One must distinguish between fixed-axis and principal-axis squeezing, between second-moment suppression and higher-order fluctuation enhancement, and between benchmarks that are mathematically convenient and those that correspond to what is actually measured in the laboratory. The central question is therefore not simply whether Janus interference modifies noise or correlations, but under which physically meaningful comparison it yields a genuine metrological advantage over the single squeezed vacuum.

The purpose of the present work is to answer that question for Janus superpositions of two single-mode squeezed vacua by focusing directly on quadrature structure and QFI. We first develop the second-moment description in a form suited to metrology. In particular, we derive explicit closed expressions for the laboratory-axis quadrature variances, for the principal squeezing selected by homodyne optimization, and for the smallest fixed-axis variance attainable when the state is optimized only within a prescribed two-state squeezed-vacuum span. This analysis already reveals a key distinction: the minimum quadrature noise of a given state and the minimum fixed-axis noise achievable within a chosen Janus span are different optimization problems, and they lead to different notions of squeezing advantage.

We then connect this quadrature analysis to explicit QFI benchmarks. Here the situation is subtler than the earlier Janus correlation studies might suggest. A state with lower laboratory-axis noise does not necessarily possess a larger QFI, and a state with strong photon-correlation signatures does not necessarily outperform the Gaussian squeezed-vacuum reference once the resource comparison is made fairly. For this reason, we analyze Janus probes under several distinct benchmarks. For number-generated phase estimation, we compare Janus states with the single squeezed vacuum at fixed mean photon number and determine when Janus interference can enhance the phase QFI. For quadratic-generator sensing, we introduce a benchmark based on fixed measured squeezing, which is especially natural experimentally because squeezing is often reported directly in dB through homodyne measurements. This allows one to ask whether two probes with the same observed squeezing can nevertheless possess substantially different metrological power.

The main conclusion is that the answer depends decisively on the benchmark. Under a fair fixed-mean-photon-number comparison, the single squeezed vacuum remains extremal for principal second-moment squeezing, so no genuine Janus advantage survives at that level. By contrast, within a prescribed two-state squeezed-vacuum span, Janus interference can reduce a fixed laboratory-axis variance below that of either constituent. The same benchmark sensitivity appears at the level of QFI. For number-generated phase estimation, a Janus state can outperform the single squeezed vacuum at fixed mean photon number only through the higher-order fluctuation structure encoded in its factorial moments. For quadratic-generator sensing, however, substantial enhancement relative to the pure-Gaussian reference can occur when the comparison is made at fixed measured squeezing. The point of the present work is therefore not merely to ask whether Janus states win or lose, but to identify precisely which notion of ``advantage'' remains meaningful under which physically motivated comparison.

This benchmark dependence is the central conceptual theme of the paper. The Janus family is analytically tractable enough to display, within a single non-Gaussian platform, how principal squeezing, fixed-axis squeezing, higher-order coherence, and QFI are connected yet fundamentally inequivalent. Because the relevant moments can be written in closed form, the family provides a transparent setting in which to separate globally fair resource statements from span-constrained interference effects. In this way, Janus states also form a useful bridge between homodyne-accessible quadrature diagnostics and the generator-based language of quantum metrology.

The paper is organized as follows. Section~\ref{sec:janus_quadrature_variances} develops the second-moment structure of the Janus family, including the laboratory-axis quadrature variances, the covariance matrix, and the principal squeezing obtained by homodyne optimization. Section~\ref{sec:janus_min_quadrature_variance} analyzes the span-constrained reduction of a fixed laboratory quadrature within a prescribed two-state squeezed-vacuum family and presents the corresponding interference landscapes. Section~\ref{sec:qfi_janus} turns these results into metrological benchmarks by evaluating the QFI for number-generated phase estimation and for quadratic-generator sensing. Section~\ref{sec:benchmark_dependence_no_go_vs_constituent} clarifies the benchmark dependence of the problem by contrasting the fixed-\(\bar n\) no-go for principal squeezing with constituent-relative compatibility of fixed-axis squeezing and phase QFI. Finally, Sec.~\ref{sec:conclusion} summarizes the main results and discusses possible directions for future work.

\section{Quadrature variances of the Janus state}
\label{sec:janus_quadrature_variances}

Quadrature fluctuations provide the most direct description of the second-moment structure of a single-mode bosonic state. In the Janus family, they are especially informative because the state is built from a coherent superposition of two squeezed vacua, so the quadrature sector shows immediately how interference redistributes noise between the laboratory axes and reshapes the fluctuation ellipse. This is also the sector accessed most directly in homodyne detection, where the measured noise depends on the phase of the local oscillator. The quadrature analysis therefore provides the natural setting in which to distinguish squeezing along a fixed laboratory axis from the principal squeezing selected intrinsically by the state.

\begin{align}
Q=\frac{a+a^\dagger}{\sqrt{2}},\qquad
P=\frac{a-a^\dagger}{i\sqrt{2}},\qquad
[Q,P]=i.
\end{align}

For the vacuum, one has \((\Delta Q)^2=(\Delta P)^2=\tfrac12\). For a general state, the second moments of \(Q\) and \(P\) are governed by the number moment \(\langle a^\dagger a\rangle\) and the anomalous moment \(\langle a^2\rangle\). Using \(a a^\dagger=a^\dagger a+1\), one finds

\begin{align}
Q^2&=\frac12\Big(a^2+a^{\dagger 2}+2a^\dagger a+1\Big),\nn\\
P^2&=\frac12\Big(-a^2-a^{\dagger 2}+2a^\dagger a+1\Big).
\end{align}

Hence

\begin{align}
\langle Q^2\rangle&=\frac12+\langle a^\dagger a\rangle+\Re\langle a^2\rangle,\nn\\
\langle P^2\rangle&=\frac12+\langle a^\dagger a\rangle-\Re\langle a^2\rangle,
\end{align}

and therefore

\begin{align}
(\Delta Q)^2&=\frac12+\langle a^\dagger a\rangle+\Re\langle a^2\rangle-\langle Q\rangle^2,\nn\\
(\Delta P)^2&=\frac12+\langle a^\dagger a\rangle-\Re\langle a^2\rangle-\langle P\rangle^2.
\label{eq:janus_fixed_axis_variances_general}
\end{align}

For the undisplaced Janus family, both constituent squeezed vacua lie in the even Fock sector, and their coherent superposition therefore also has even parity. It follows that \(\langle a\rangle=0\), so \(\langle Q\rangle=\langle P\rangle=0\) \cite{Azizi2026Janus,Azizi2025Janus_higher}. In that case, the fixed-axis quadrature variances are determined entirely by \(\langle a^\dagger a\rangle\) and \(\Re\langle a^2\rangle\).

The Janus state is the coherent superposition

\begin{align}
\ket{\psi}=\chi\ket{\xi}+\eta\ket{\zeta},
\end{align}

with \(\chi,\eta\in\mathbb{C}\) and normalization

\begin{align}
\braket{\psi}{\psi}=|\chi|^2+|\eta|^2
+2\,\Re\Big[\chi\eta^\ast\braket{\zeta}{\xi}\Big]
=1.
\label{eq:janus_norm_constraint}
\end{align}

The two constituent squeezed vacua are parameterized as

\begin{align}
\xi=r e^{i\theta},\qquad
\zeta=s e^{i\phi},
\end{align}

with \(r,s\ge 0\) and \(\theta,\phi\in\mathbb{R}\). It is convenient to introduce

\begin{align}
\alpha=\tanh r\,e^{i\theta},\qquad
\beta=\tanh s\,e^{i\phi}.
\end{align}

We then define

\begin{align}
x\equiv |\alpha|^2=\tanh^2 r,\qquad
y\equiv |\beta|^2=\tanh^2 s,
\end{align}

and

\begin{align}
z\equiv& \alpha\beta^\ast
=\tanh r\,\tanh s\,e^{i(\theta-\phi)}
=\sqrt{xy}\,e^{i\Delta},
\nn\\
\Delta\equiv&\theta-\phi.
\end{align}

Since \(0\le x,y<1\), one has \(|\alpha|<1\), \(|\beta|<1\), and \(|z|=\sqrt{xy}<1\). Accordingly, the factors \((1-z)^{-1/2}\) and \((1-z)^{-3/2}\) are well defined on the principal branch.

In disentangled form,

\begin{align}
\ket{\xi}
&=(1-x)^{1/4}\exp\Big(-\frac{\alpha}{2}a^{\dagger 2}\Big)\ket{0},\nn\\
\bra{\zeta}
&=(1-y)^{1/4}\bra{0}\exp\Big(-\frac{\beta^\ast}{2}a^{2}\Big),
\end{align}

and therefore the overlap is \cite{Azizi2026Janus,Azizi2025Janus_higher}

\begin{align}
\braket{\zeta}{\xi}
=(1-x)^{1/4}(1-y)^{1/4}(1-z)^{-1/2}.
\label{eq:sq_overlap_closed}
\end{align}

With this notation, the expectation value of any operator \(O\) decomposes as

\begin{align}
\langle O\rangle
=&|\chi|^2\bra{\xi}O\ket{\xi}
+|\eta|^2\bra{\zeta}O\ket{\zeta}\nn\\
&+\chi^\ast\eta\,\bra{\xi}O\ket{\zeta}
+\chi\eta^\ast\,\bra{\zeta}O\ket{\xi},
\end{align}

with \(\chi\) and \(\eta\) constrained by Eq.~\eqref{eq:janus_norm_constraint}.

We first evaluate \(\langle a^\dagger a\rangle\). For a single squeezed vacuum, \(\langle a^\dagger a\rangle=\sinh^2 r=x/(1-x)\), and similarly \(\langle a^\dagger a\rangle=y/(1-y)\) in \(\ket{\zeta}\). The cross matrix elements are \cite{Azizi2026Janus,Azizi2025Janus_higher}

\begin{align}
\bra{\zeta}a^\dagger a\ket{\xi}
&=(1-x)^{1/4}(1-y)^{1/4}\,\frac{z}{(1-z)^{3/2}},\nn\\
\bra{\xi}a^\dagger a\ket{\zeta}
&=(1-x)^{1/4}(1-y)^{1/4}\,\frac{z^\ast}{(1-z^\ast)^{3/2}}.
\end{align}

Substituting these contributions into the general decomposition gives

\begin{align}
\langle a^\dagger a\rangle
=&\ |\chi|^2\frac{x}{1-x}
+|\eta|^2\frac{y}{1-y} \label{eq:janus_nbar}\\
&+2\,\Re\Bigg[
\chi\eta^\ast(1-x)^{1/4}(1-y)^{1/4}\,
\frac{z}{(1-z)^{3/2}}
\Bigg]. \nn
\end{align}

We next evaluate the anomalous moment \(\langle a^2\rangle\). For the diagonal terms,

\begin{align}
\bra{\xi}a^2\ket{\xi}
=-e^{i\theta}\sinh r\cosh r
=-\frac{\alpha}{1-x},
\end{align}

and similarly

\begin{align}
\bra{\zeta}a^2\ket{\zeta}
=-\frac{\beta}{1-y}.
\end{align}

The remaining ingredient is the cross moment \(\bra{\zeta}a^2\ket{\xi}\). This is the \(k=1\) special case of the closed-form expression for \(\bra{\zeta}a^{2k}\ket{\xi}\) derived in Appendix~\ref{app:cross_even_moments}. Setting \(k=1\) gives

\begin{align}
\bra{\zeta}a^2\ket{\xi}
&=-(1-x)^{1/4}(1-y)^{1/4}\,\tanh r\,e^{i\theta}\,\frac{1}{(1-z)^{3/2}},\nn\\
\bra{\xi}a^2\ket{\zeta}
&=-(1-x)^{1/4}(1-y)^{1/4}\,\tanh s\,e^{i\phi}\,\frac{1}{(1-z^\ast)^{3/2}}.
\end{align}

Combining the diagonal and interference terms then yields

\begin{align}
\langle a^2\rangle
=-\Bigg\{
&|\chi|^2\frac{\tanh r\,e^{i\theta}}{1-x}
+|\eta|^2\frac{\tanh s\,e^{i\phi}}{1-y} \label{eq:janus_a2}\\
&+\chi\eta^\ast(1-x)^{1/4}(1-y)^{1/4}\,
\frac{\tanh r\,e^{i\theta}}{(1-z)^{3/2}}\nn\\
&+\chi^\ast\eta(1-x)^{1/4}(1-y)^{1/4}\,
\frac{\tanh s\,e^{i\phi}}{(1-z^\ast)^{3/2}}
\Bigg\}.
\nn
\end{align}

For the undisplaced Janus family, the fixed-axis quadrature variances therefore take the form

\begin{align}
(\Delta Q)^2&=\frac12+\langle a^\dagger a\rangle+\Re\langle a^2\rangle,\nn\\
(\Delta P)^2&=\frac12+\langle a^\dagger a\rangle-\Re\langle a^2\rangle,
\label{eq:janus_fixed_axis_variances_undisplaced}
\end{align}

with \(\langle a^\dagger a\rangle\) and \(\langle a^2\rangle\) given explicitly by Eqs.~\eqref{eq:janus_nbar} and \eqref{eq:janus_a2}.

\subsection{Homodyne-accessible quadratures and principal squeezing}

The fixed laboratory quadratures \(Q\) and \(P\) provide an important first description of the noise, but in general they do not coincide with the directions naturally selected by the state. In experiment, quadrature fluctuations are measured by balanced homodyne detection, where the phase of the local oscillator determines which rotated quadrature is probed. For that reason, it is not sufficient to examine only the special choices \(Q\) and \(P\). One must also identify the quadrature direction along which a given Janus state attains its smallest second-moment noise. This optimization is governed entirely by the covariance matrix and therefore admits a closed algebraic solution once \(\langle a^\dagger a\rangle\) and \(\langle a^2\rangle\) are known.

To place the fixed-axis and rotated-axis analyses on the same footing, it is convenient to introduce the displacement-invariant central moments

\begin{align}
\bar n\equiv \langle a^\dagger a\rangle-|\langle a\rangle|^2,
\qquad
m\equiv \langle a^2\rangle-\langle a\rangle^2.
\label{eq:nbar_m_def}
\end{align}

For the undisplaced even-parity Janus family, \(\langle a\rangle=0\), and therefore \(\bar n=\langle a^\dagger a\rangle\) and \(m=\langle a^2\rangle\). In terms of \(\bar n\) and \(m\), the fixed-axis variances become

\begin{align}
(\Delta Q)^2&=\frac12+\bar n+\Re m,\nn\\
(\Delta P)^2&=\frac12+\bar n-\Re m.
\label{eq:fixed_axis_variances_nbar_m}
\end{align}

The symmetrized covariance is

\begin{align}
\frac12\langle (Q-\langle Q\rangle)(P-\langle P\rangle)
+(P-\langle P\rangle)(Q-\langle Q\rangle)\rangle
=\Im m,
\label{eq:qp_covariance_janus}
\end{align}

so the covariance matrix may be written as

\begin{align}
V_{QQ}&=\frac12+\bar n+\Re m,\nn\\
V_{PP}&=\frac12+\bar n-\Re m,\nn\\
V_{QP}&=V_{PQ}=\Im m,
\end{align}

or equivalently

\begin{align}
\mathbf{V}=
\begin{pmatrix}
\frac12+\bar n+\Re m & \Im m\\
\Im m & \frac12+\bar n-\Re m
\end{pmatrix}.
\label{eq:covariance_matrix_Janus}
\end{align}

The rotated quadrature selected by a homodyne phase \(\varphi\) is

\begin{align}
X_\varphi
=Q\cos\varphi+P\sin\varphi
=\frac{a e^{-i\varphi}+a^\dagger e^{i\varphi}}{\sqrt{2}}.
\end{align}

Its variance is

\begin{align}
(\Delta X_\varphi)^2
=\frac12+\bar n+\Re\Big(m\,e^{-2i\varphi}\Big).
\label{eq:var_rotated_quadrature_compact}
\end{align}

Accordingly, the minimum and maximum second-moment noises over all homodyne phases are the eigenvalues of \(\mathbf{V}\),

\begin{align}
(\Delta X)^2_{\min}
=\frac12+\bar n-|m|,
\qquad
(\Delta X)^2_{\max}
=\frac12+\bar n+|m|.
\label{eq:principal_variances_in_terms_of_nbar_m}
\end{align}

The principal minimum always satisfies \((\Delta X)^2_{\min}\le \min\{(\Delta Q)^2,(\Delta P)^2\}\), with equality only when the squeezing axis is already aligned with one of the laboratory axes.

The minimizing homodyne phase \(\varphi_\ast\) is characterized by \(\Re(m e^{-2i\varphi_\ast})=-|m|\), equivalently

\begin{align}
e^{-2i\varphi_\ast}=-\,\frac{m^\ast}{|m|}
\qquad\text{or}\qquad
e^{2i\varphi_\ast}=-\,\frac{m}{|m|},
\label{eq:theta_star_phase_condition}
\end{align}

which implies

\begin{align}
\varphi_\ast=\frac12\arg(m)+\frac{\pi}{2}
\qquad (\mathrm{mod}\ \pi).
\label{eq:theta_star_argm}
\end{align}

For the undisplaced Janus family, the principal-axis minimum therefore reduces to

\begin{align}
(\Delta X)^2_{\min}
=\frac12+\langle a^\dagger a\rangle-|\langle a^2\rangle|.
\label{eq:Janus_principal_min_variance_closed}
\end{align}

This expression is especially useful because it separates two distinct physical contributions. The moment \(\langle a^\dagger a\rangle\) sets the overall scale of fluctuations, while \(|\langle a^2\rangle|\) quantifies how efficiently those fluctuations are redistributed into a squeezed and an anti-squeezed quadrature pair. For Janus states, both quantities depend on the same interference structure encoded in the coefficients \(\chi,\eta\) and in the overlap parameter \(z\), so principal squeezing provides a direct homodyne-accessible measure of coherent Janus interference.

The displaced Janus family is obtained by applying a common displacement to the entire superposition,

\begin{align}
\ket{\psi_{\mathrm{DJ}}}=D(\alpha_0)\ket{\psi},
\end{align}

where \(\alpha_0\) denotes the displacement amplitude. Using \(D^\dagger(\alpha_0)\,a\,D(\alpha_0)=a+\alpha_0\), one finds

\begin{align}
\langle a^\dagger a\rangle_{\mathrm{DJ}}
&=\langle a^\dagger a\rangle_{\psi}+|\alpha_0|^2,\nn\\
\langle a^2\rangle_{\mathrm{DJ}}
&=\langle a^2\rangle_{\psi}+\alpha_0^2,\nn\\
\langle a\rangle_{\mathrm{DJ}}
&=\alpha_0,
\end{align}

since \(\langle a\rangle_\psi=0\) for the undisplaced even-parity Janus state. It follows immediately that the central moments are unchanged,

\begin{align}
\bar n_{\mathrm{DJ}}=\bar n_\psi,\qquad
m_{\mathrm{DJ}}=m_\psi.
\end{align}

Hence the covariance matrix \eqref{eq:covariance_matrix_Janus}, the rotated-quadrature variance \eqref{eq:var_rotated_quadrature_compact}, the principal variances \eqref{eq:principal_variances_in_terms_of_nbar_m}, and the optimal-axis condition \eqref{eq:theta_star_argm} all apply unchanged to the commonly displaced family. In particular,

\begin{align}
(\Delta Q)^2_{\mathrm{DJ}}=(\Delta Q)^2_{\psi},\qquad
(\Delta P)^2_{\mathrm{DJ}}=(\Delta P)^2_{\psi},
\end{align}

and likewise the principal minimum and maximum variances are displacement invariant.

Accordingly, the second-moment squeezing properties of the Janus family are completely determined by the central moments \(\bar n\) and \(m\). These two quantities fix the laboratory-axis variances, the rotation of the noise ellipse, and the principal squeezing accessible in a homodyne scan of the local-oscillator phase. They are therefore the natural variables for the second-moment and metrological benchmarks developed below.

\section{Regime of minimal quadrature variance in the Janus family}
\label{sec:janus_min_quadrature_variance}

This section addresses an optimization distinct from the principal-axis homodyne problem of Sec.~\ref{sec:janus_quadrature_variances}. There the Janus state was fixed and the homodyne phase \(\varphi\) was varied, leading to the principal-axis quantity \((\Delta X)^2_{\min}\). Here the laboratory quadrature is fixed, the two constituent squeezed vacua are fixed, and only the Janus superposition coefficients are varied, leading instead to a span-constrained quantity such as \((\Delta Q)^2_{\min}\). For the undisplaced even-parity Janus family, parity implies \(\langle a\rangle=0\), and therefore

\begin{align}
(\Delta Q)^2=\langle Q^2\rangle,\qquad
(\Delta P)^2=\langle P^2\rangle.
\end{align}

The task is thus to minimize \(\langle Q^2\rangle\) or \(\langle P^2\rangle\) inside the normalized two-state span generated by \(\ket{\xi}\) and \(\ket{\zeta}\). This is a span-constrained optimization, not a global minimization over all bosonic states at fixed resource.

\subsection{Generalized eigenvalue formulation}

Let

\begin{align}
\mathcal{S}\equiv\braket{\zeta}{\xi}
\end{align}

denote the overlap of the two squeezed vacua. The Gram matrix of the nonorthogonal basis is

\begin{align}
\mathbf{S}\equiv
\begin{pmatrix}
\braket{\xi}{\xi} & \braket{\xi}{\zeta}\\
\braket{\zeta}{\xi} & \braket{\zeta}{\zeta}
\end{pmatrix}
=
\begin{pmatrix}
1 & \mathcal{S}^\ast\\
\mathcal{S} & 1
\end{pmatrix},
\end{align}

while the matrix representation of \(Q^2\) in the same span is

\begin{align}
\mathbf{M}_Q\equiv
\begin{pmatrix}
\bra{\xi}Q^2\ket{\xi} & \bra{\xi}Q^2\ket{\zeta}\\
\bra{\zeta}Q^2\ket{\xi} & \bra{\zeta}Q^2\ket{\zeta}
\end{pmatrix}
=
\begin{pmatrix}
A_Q & C_Q^\ast\\
C_Q & B_Q
\end{pmatrix},
\end{align}

with

\begin{align}
A_Q=&\bra{\xi}Q^2\ket{\xi},\qquad
B_Q=\bra{\zeta}Q^2\ket{\zeta},\nn\\
C_Q=&\bra{\zeta}Q^2\ket{\xi}.
\end{align}

For a coefficient vector \(c=(\chi,\eta)^T\), the fixed-axis variance inside this span is the Rayleigh quotient

\begin{align}
(\Delta Q)^2=\frac{c^\dagger \mathbf{M}_Q c}{c^\dagger \mathbf{S}\,c}.
\end{align}

Minimizing \((\Delta Q)^2\) subject to the normalization constraint \(c^\dagger \mathbf{S}c=1\) yields the generalized eigenvalue problem

\begin{align}
\mathbf{M}_Q\,c=\lambda_Q\,\mathbf{S}\,c,
\label{eq:gen_eig_Q2}
\end{align}

and the smaller generalized eigenvalue gives the minimum attainable fixed-axis noise,

\begin{align}
(\Delta Q)^2_{\min}=\lambda_{Q,-}.
\end{align}

When \(|\mathcal{S}|<1\), the Gram matrix \(\mathbf{S}\) is positive definite and the span is genuinely two-dimensional. In that case the generalized eigenvalues are real. As \(|\mathcal{S}|\to 1\), for example when the two squeezed vacua approach one another, the span collapses to one dimension and the optimization reduces to the corresponding single-state value. Numerically, that near-singular regime is better treated after orthonormalizing the span, for example by a Cholesky reduction of \(\mathbf{S}\), before diagonalizing the transformed operator.

Because the problem is \(2\times2\), the generalized spectrum is available in closed form. The characteristic equation \(\det(\mathbf{M}_Q-\lambda\,\mathbf{S})=0\) becomes

\begin{align}
0
=&\,(1-|\mathcal{S}|^2)\lambda^2 \nn\\
&-\Big[(A_Q+B_Q)-2\,\Re\big(\mathcal{S}^\ast C_Q\big)\Big]\lambda \nn\\
&+\big(A_QB_Q-|C_Q|^2\big),
\label{eq:charpoly_Q}
\end{align}

so that

\begin{align}
\lambda_{Q,\pm}
=&\,\frac{(A_Q+B_Q)-2\,\Re\big(\mathcal{S}^\ast C_Q\big)}
{2(1-|\mathcal{S}|^2)}
\pm \frac{1}{2(1-|\mathcal{S}|^2)}\sqrt{\mathcal{D}_Q},
\label{eq:lambda_Q_pm}
\end{align}

with

\begin{align}
\mathcal{D}_Q
=&\,\Big[(A_Q+B_Q)-2\,\Re\big(\mathcal{S}^\ast C_Q\big)\Big]^2 \nn\\
&-4(1-|\mathcal{S}|^2)\big(A_QB_Q-|C_Q|^2\big).
\end{align}

Since \(\mathbf{M}_Q\) is Hermitian and \(\mathbf{S}\) is positive definite, one has \(\mathcal{D}_Q\ge 0\).

The minimizing generalized eigenvector fixes the optimal coefficient ratio. Using the first row of Eq.~\eqref{eq:gen_eig_Q2}, one finds

\begin{align}
\frac{\eta}{\chi}
=-\,\frac{A_Q-\lambda_{Q,-}}
{C_Q^\ast-\lambda_{Q,-}\mathcal{S}^\ast}.
\label{eq:opt_ratio_Q}
\end{align}

The construction for \(P^2\) is completely analogous: one replaces \(Q^2\) by \(P^2\), constructs \(\mathbf{M}_P\), and solves the same generalized eigenvalue problem.

\subsection{\texorpdfstring{Closed matrix elements in terms of \((r,s,\theta,\phi)\) and \((x,y,z)\)}{}}

The matrix elements entering the generalized-eigenvalue problem may be written explicitly in the invariant notation introduced in Sec.~\ref{sec:janus_quadrature_variances},

\begin{align}
x=&\tanh^2 r,\qquad
y=\tanh^2 s,\nn\\
z=&\tanh r\,\tanh s\,e^{i(\theta-\phi)}
=\sqrt{xy}\,e^{i\Delta},
\end{align}

with \(\Delta=\theta-\phi\). In these variables, the overlap is \cite{Azizi2026Janus,Azizi2025Janus_higher}

\begin{align}
\mathcal{S}
=(1-x)^{1/4}(1-y)^{1/4}\,\frac{1}{(1-z)^{1/2}}.
\label{eq:overlap_S_min}
\end{align}

The diagonal entries follow from the single-squeezed-vacuum moments. In \(\ket{\xi}\), \(\langle a^\dagger a\rangle=x/(1-x)\) and \(\langle a^2\rangle=-(\tanh r\,e^{i\theta})/(1-x)\), with the corresponding expressions for \(\ket{\zeta}\) obtained by \((x,\theta)\to(y,\phi)\). Using \(\bra{\xi}Q^2\ket{\xi}=\tfrac12+\langle a^\dagger a\rangle+\Re\langle a^2\rangle\), one obtains

\begin{align}
A_Q
&=\frac12+\frac{x}{1-x}-\Re\Big(\frac{\sqrt{x}\,e^{i\theta}}{1-x}\Big)
=\frac{1+x-2\sqrt{x}\cos\theta}{2(1-x)},\nn\\
B_Q
&=\frac12+\frac{y}{1-y}-\Re\Big(\frac{\sqrt{y}\,e^{i\phi}}{1-y}\Big)
=\frac{1+y-2\sqrt{y}\cos\phi}{2(1-y)}.
\end{align}

The off-diagonal element \(C_Q=\bra{\zeta}Q^2\ket{\xi}\) follows from \(Q^2=\tfrac12(a^2+a^{\dagger 2}+2a^\dagger a+1)\) together with the closed cross moments \cite{Azizi2026Janus,Azizi2025Janus_higher}:

\begin{align}
\bra{\zeta}a^\dagger a\ket{\xi}
&=(1-x)^{1/4}(1-y)^{1/4}\,\frac{z}{(1-z)^{3/2}},\nn\\
\bra{\zeta}a^{2}\ket{\xi}
&=(1-x)^{1/4}(1-y)^{1/4}\,
\frac{-\sqrt{x}\,e^{i\theta}}{(1-z)^{3/2}},\nn\\
\bra{\zeta}a^{\dagger 2}\ket{\xi}
&=(1-x)^{1/4}(1-y)^{1/4}\,
\frac{-\sqrt{y}\,e^{-i\phi}}{(1-z)^{3/2}}.
\end{align}

Combining these terms gives

\begin{align}
C_Q
=&(1-x)^{1/4}(1-y)^{1/4}\,
\frac{1+z-\sqrt{x}\,e^{i\theta}-\sqrt{y}\,e^{-i\phi}}{2(1-z)^{3/2}}.
\label{eq:CQ_compact}
\end{align}

For generic complex \(z\), the characteristic coefficients in Eq.~\eqref{eq:charpoly_Q} depend on \(|\mathcal{S}|^2\), \(\Re(\mathcal{S}^\ast C_Q)\), and \(|C_Q|^2\), so the phase of \(z\) must be retained until those combinations are formed. The equal-strength reduction below evaluates these combinations explicitly on the slice of principal interest.

Although the present optimization focuses on \((\Delta Q)^2\), it is useful to record the corresponding \(P^2\) matrix element because later coefficient-space comparisons between squeezing and anti-squeezing involve both laboratory axes. Since \(P^2=\tfrac12(-a^2-a^{\dagger 2}+2a^\dagger a+1)\), one finds

\begin{align}
C_P
=&(1-x)^{1/4}(1-y)^{1/4}\,
\frac{1+z+\sqrt{x}\,e^{i\theta}+\sqrt{y}\,e^{-i\phi}}{2(1-z)^{3/2}}.
\label{eq:CP_compact}
\end{align}

\subsection{\texorpdfstring{The regime that minimizes \((\Delta Q)^2\)}{}}

The minimum found above must be interpreted with care. The squeezing parameters \((r,s,\theta,\phi)\) fix the two constituent vacua, and only the coefficients \((\chi,\eta)\) are varied. Accordingly, \((\Delta Q)^2_{\min}\) may lie below the smaller constituent value \(\min(A_Q,B_Q)\), but that reduction is typically accompanied by enlarged fluctuations in the conjugate quadrature and often by an increased photon number. A lower fixed-axis variance produced by coefficient interference therefore does not by itself imply a metrological advantage; such a claim becomes meaningful only after imposing an explicit benchmark, such as fixed \(\langle a^\dagger a\rangle\), fixed squeezing strength, or a bound on the uncertainty product.

The generalized-eigenvalue formulation also makes precise when interference can or cannot modify the coefficient-optimized variance. If \(\mathbf{M}_Q\) is proportional to the overlap matrix \(\mathbf{S}\), then \(\lambda_{Q,+}=\lambda_{Q,-}\) and every normalized state in the span has the same \((\Delta Q)^2\). In that case no superposition can outperform the constituent squeezed vacua. More generally, interference changes the coefficient-optimized fixed-axis variance whenever \(\mathbf{M}_Q\) is not proportional to \(\mathbf{S}\). When \(\mathcal{S}\neq 0\), exact proportionality requires \(A_Q=B_Q=C_Q/\mathcal{S}\) with a common real value.

A physically transparent regime in which this misalignment is especially pronounced is obtained when the two squeezing axes are nearly aligned. Taking \(\Delta=\theta-\phi\simeq 0\), and in particular \(\theta\simeq\phi\simeq 0\), places \(Q\) close to the squeezed quadrature of each constituent. Then \(z=\sqrt{xy}\,e^{i\Delta}\) is real and positive to good approximation, and the kernels \((1-z)^{-1/2}\) and \((1-z)^{-3/2}\) are enhanced as \(z\to 1\). This is precisely the regime in which the off-diagonal structure entering the generalized-eigenvalue problem becomes most pronounced.

A second important ingredient is near-equal squeezing strength, \(r\simeq s\), while keeping the two states distinct. Then \(|\mathcal{S}|\) may approach unity, so the normalization constraint \(c^\dagger\mathbf{S}c=1\) strongly couples the two coefficient channels and allows large constructive or destructive interference. The same regime is numerically delicate because \(\mathbf{S}\) becomes nearly singular, and the explicit quadratic formulas may lose significance. Stable generalized eigensolvers or an explicit orthonormalization of the span are therefore preferable when \(|\mathcal{S}|\approx 1\).

In this aligned, near-overlap regime, the minimizing state typically has subtractive character: the optimal generalized eigenvector often carries a relative minus sign between its two components, equivalently a relative coefficient phase near \(\pi\). In that way, the dominant overlap-driven contributions to \(\langle Q^2\rangle\) can partially cancel while normalization is still enforced through \(\mathbf{S}\). Quantitatively, the coefficient ratio is fixed by Eq.~\eqref{eq:opt_ratio_Q} evaluated at \(\lambda_{Q,-}\).

As a representative numerical example, take \(\theta=\phi=0\) and two moderately close squeezing strengths, for instance \(r=1\) and \(s=0.9\). Evaluating \(\lambda_{Q,-}\) with a stable generalized eigensolver applied to \((\mathbf{M}_Q,\mathbf{S})\) gives

\begin{align}
(\Delta Q)^2_{\min}\approx 4.09\times10^{-2},\qquad
\frac{\eta}{\chi}\approx -0.80,
\end{align}

which is smaller than the better constituent squeezed-vacuum value, \((\Delta Q)^2_{\xi}=\tfrac12 e^{-2r}\approx 6.77\times10^{-2}\). The conjugate quadrature grows accordingly, so the uncertainty relation remains intact. This example illustrates the basic mechanism already identified by the generalized-eigenvalue analysis: reduced \(Q\)-noise inside the fixed span is purchased by larger fluctuations elsewhere, which is why later metrological comparisons must be made against an explicit resource benchmark.

\subsection{\texorpdfstring{Equal-strength family \(r=s\): analytic minimum of \((\Delta Q)^2\)}{}}
\label{sec:equal_strength_min_DQ}

The equal-strength family \(r=s\) provides a particularly useful reduction for which the generalized minimum can be written analytically. In this case, \(x=y=\tanh^2 r\) and \(z=\tanh^2 r\,e^{i\Delta}=x\,e^{i\Delta}\), with \(\Delta\equiv\theta-\phi\). As long as \(\Delta\not\equiv 0 \pmod{2\pi}\), the two squeezed vacua remain linearly independent, \(\mathbf{S}\) is nonsingular, and the generalized-eigenvalue minimum \((\Delta Q)^2_{\min}=\lambda_{Q,-}\) from Eq.~\eqref{eq:lambda_Q_pm} is well defined. The specialization is obtained by evaluating the coefficients of Eq.~\eqref{eq:charpoly_Q} at \(r=s\) and substituting them into Eq.~\eqref{eq:lambda_Q_pm}; the intermediate algebra is collected in Appendix~\ref{app:equal_strength_min_DQ}.

In the equal-strength family, the overlap and matrix elements reduce to

\begin{align}
\mathcal{S}
=(1-x)^{1/2}\,(1-xe^{i\Delta})^{-1/2},
\qquad
|\mathcal{S}|^2=\frac{1-x}{|1-xe^{i\Delta}|},
\label{eq:S_equal_strength_main}
\end{align}

\begin{align}
A_Q
=\frac{1+x-2\sqrt{x}\cos\theta}{2(1-x)},
\qquad
B_Q
=\frac{1+x-2\sqrt{x}\cos\phi}{2(1-x)},
\label{eq:AB_equal_strength_main}
\end{align}

\begin{align}
C_Q
=&(1-x)^{1/2}\,
\frac{\mathcal{N}_Q}{2(1-xe^{i\Delta})^{3/2}},
\nn\\
\mathcal{N}_Q\equiv & 1+xe^{i\Delta}-\sqrt{x}\,e^{i\theta}-\sqrt{x}\,e^{-i\phi}.
\label{eq:CQ_equal_strength_main}
\end{align}

The \(\Delta\)-dependence enters through \(|1-xe^{i\Delta}|=\sqrt{1+x^2-2x\cos\Delta}\) together with the phase relation \(\Delta=\theta-\phi\). For \(\Delta\neq 0\), the coefficients appearing in Eq.~\eqref{eq:charpoly_Q} become

\begin{align}
\alpha
=&(A_Q+B_Q) \nn\\
&-\frac{(1-x)^2}{|1-xe^{i\Delta}|^{3}}
\Big[(1+x)-\sqrt{x}\,(\cos\theta+\cos\phi)\Big],\nn\\
\beta
=&A_QB_Q
-\frac{(1-x)}{4\,|1-xe^{i\Delta}|^{3}}\,
|\mathcal{N}_Q|^2,\nn\\
\gamma
=&1-\frac{1-x}{|1-xe^{i\Delta}|},
\label{eq:alpha_beta_gamma_equal_strength_main}
\end{align}

where \(|\mathcal{N}_Q|^2\) is given explicitly in Appendix~\ref{app:equal_strength_min_DQ}. Substituting these coefficients into Eq.~\eqref{eq:lambda_Q_pm} yields

\begin{align}
(\Delta Q)^2_{\min}=\lambda_{Q,-}
=\frac{\alpha-\sqrt{\alpha^2-4\gamma\beta}}{2\gamma},
\qquad
(\Delta\neq 0),
\label{eq:DQmin_equal_strength_final}
\end{align}

with \(\Delta Q_{\min}=\sqrt{\lambda_{Q,-}}\).

\begin{figure}[t]
\centering
\includegraphics[width=\columnwidth]{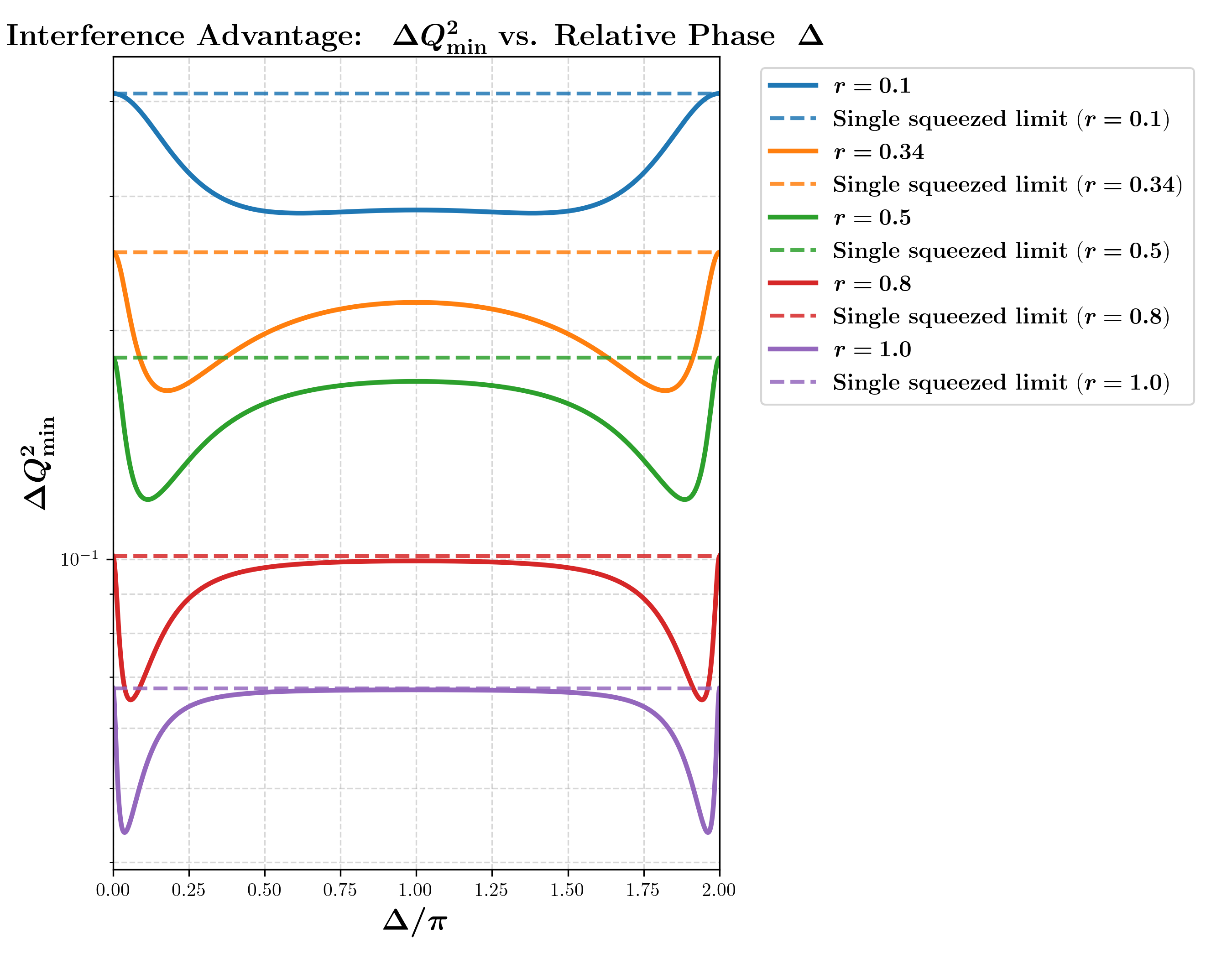}
\caption{Equal-strength (\(r=s\)) Janus minimum quadrature noise \((\Delta Q)^2_{\min}\) versus the relative squeezing phase \(\Delta=\theta-\phi\), computed from Eq.~\eqref{eq:DQmin_equal_strength_final} for \(\Delta\neq 0\) on the slice \(\theta=0\) and \(\phi=-\Delta\). Solid curves show the optimized two-state-span minimum, while dashed curves show the single squeezed-vacuum benchmark at the same strength, \((\Delta Q)^2=\tfrac12 e^{-2r}\). The curves are rendered continuous at \(\Delta=0\) and \(2\pi\) by connecting to the one-dimensional identical-state value.}
\label{fig:interference_advantage_5_rs_regularized}
\end{figure}

\begin{figure}[t]
\centering
\includegraphics[width=\columnwidth]{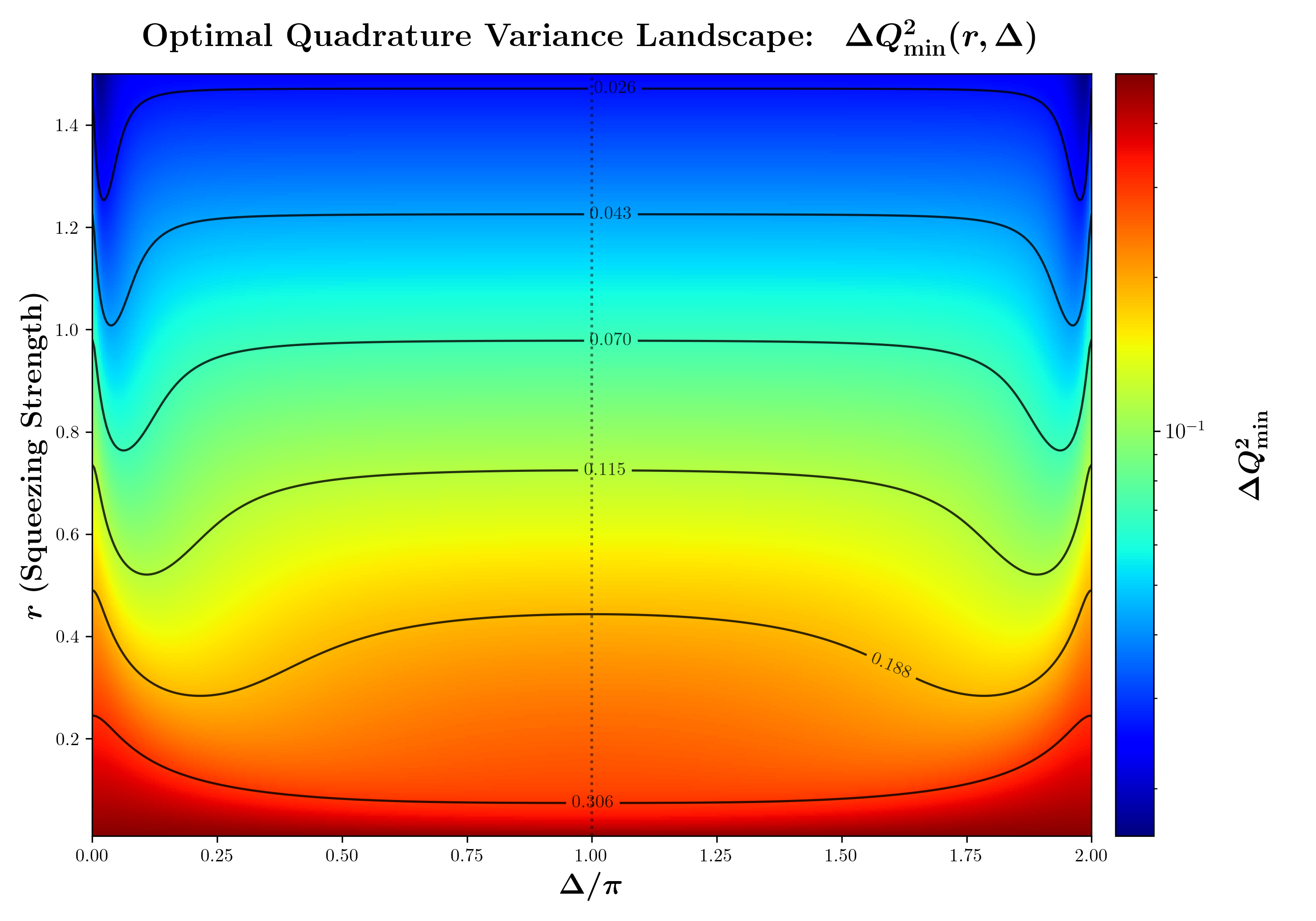}
\caption{Two-parameter landscape of the equal-strength Janus optimum \((\Delta Q)^2_{\min}(r,\Delta)\) for the slice \(\theta=0\) and \(\phi=-\Delta\). Colors show \((\Delta Q)^2_{\min}\) on a logarithmic scale, with labeled contour lines indicating constant-variance levels. The vertical guide at \(\Delta=\pi\) marks the maximally phase-opposed configuration and organizes a broad region of interference-enhanced squeezing.}
\label{fig:interference_landscape_2d_heatmap}
\end{figure}

Figure~\ref{fig:interference_advantage_5_rs_regularized} shows a natural experimental slice in which one constituent is aligned with \(Q\) by setting \(\theta=0\), while the relative phase is scanned through \(\phi=-\Delta\). The dashed curves indicate the single-squeezed-vacuum benchmark at the same strength, \((\Delta Q)^2=\tfrac12 e^{-2r}\). In this plotted slice, the optimized two-state-span minimum remains below that benchmark throughout the interior \(0<\Delta<2\pi\), and approaches it as \(\Delta\to 0\) or \(2\pi\), where the two squeezed vacua become identical and the span collapses. Figure~\ref{fig:interference_landscape_2d_heatmap} displays the corresponding two-parameter landscape \((\Delta Q)^2_{\min}(r,\Delta)\) for the same slice, showing a broad region of interference-enhanced squeezing and the organizing role of the line \(\Delta=\pi\).

The identical-state limit is recovered by taking \(\Delta\to 0\). Then \(|\mathcal{S}|\to 1\), the two-dimensional span becomes one-dimensional, and the generalized minimum reduces to the ordinary squeezed-vacuum variance:

\begin{align}
(\Delta Q)^2_{\min}\to A_Q
=&\frac12\Big(\cosh 2r-\sinh 2r \cos\theta\Big)\nn\\
=&\frac12 e^{-2r}+\sinh(2r)\,\sin^2\Big(\frac{\theta}{2}\Big).
\label{eq:DQmin_identical_limit_rform}
\end{align}

\subsection{\texorpdfstring{Numerical landscapes of $\Delta Q$, $\Delta P$, and $\Delta Q\,\Delta P$}{}}
\label{subsec:janus_plots_sweet_spots}

This part turns the closed second-moment formulas into coefficient-space landscapes. The purpose is to visualize how Janus interference reshapes the laboratory-axis cross sections of the quadrature-noise ellipse across the admissible coefficient space and to relate that second-moment structure to intensity fluctuations. Throughout, the constituent squeezed vacua are held fixed within each panel, while the superposition coefficients are varied subject to the state-normalization constraint in Eq.~\eqref{eq:janus_norm_constraint}.

In this section we display the quadrature \emph{standard deviations} \(\Delta Q=\sqrt{\langle (Q-\langle Q\rangle)^2\rangle}\) and \(\Delta P=\sqrt{\langle (P-\langle P\rangle)^2\rangle}\), whose vacuum benchmark is \(1/\sqrt{2}\). This differs only in presentation from Secs.~\ref{sec:janus_quadrature_variances}--\ref{sec:janus_min_quadrature_variance}, where the discussion was written primarily in terms of the corresponding variances \((\Delta Q)^2\) and \((\Delta P)^2\), with vacuum benchmark \(1/2\).

To scan the coefficient space without introducing an additional normalization prefactor, the global phase is fixed by choosing \(\chi\in\mathbb{R}_{\ge 0}\) and writing

\begin{align}
\eta=|\eta|e^{i\delta},\qquad \delta\in[0,2\pi).
\label{eq:eta_polar_param}
\end{align}

For each pair \((|\eta|,\delta)\), the normalization condition becomes a quadratic equation for \(\chi\). Only points for which this equation admits a real nonnegative solution are retained in the plots, so each heat map is understood as restricted to the physically admissible coefficient region. Within that admissible region, one branch is chosen consistently after fixing the global phase by \(\chi\in\mathbb{R}_{\ge 0}\). In all heat maps, the horizontal axis is \(\delta/\pi\in[0,2]\), the vertical axis is the relative amplitude \(|\eta|\), and the relative squeezing-angle offset between the two constituents is denoted by \(\Delta=\theta-\phi\).

We begin with the fixed laboratory quadratures. The landscapes \(\Delta Q(|\eta|,\delta)\) and \(\Delta P(|\eta|,\delta)\) show directly where interference lowers the noise along one chosen axis and where it necessarily increases the conjugate noise. The dashed contour labeled ``Vac'' marks the vacuum benchmark \(\Delta Q=1/\sqrt{2}\) or \(\Delta P=1/\sqrt{2}\), so regions inside that contour correspond to laboratory-axis squeezing. These maps should therefore be read together: a basin of reduced \(\Delta Q\) typically comes with a compensating increase in \(\Delta P\), reflecting a redistribution of fluctuations rather than a uniform suppression of noise.

A compact summary of this trade-off is the laboratory-axis uncertainty product \(\Delta Q\,\Delta P\), which obeys the Heisenberg bound \(\Delta Q\,\Delta P\ge 1/2\). Mapping \(\Delta Q\,\Delta P\) over the coefficient space reveals where the state approaches that lower bound in the fixed laboratory frame and how sharply the near-minimum-uncertainty regions are localized. When \(Q\)--\(P\) correlations are appreciable, the sharper statement is the Robertson--Schr\"odinger relation \((\Delta Q)^2(\Delta P)^2-\mathrm{Cov}(Q,P)^2\ge 1/4\), with \(\mathrm{Cov}(Q,P)=\tfrac12\langle \{Q-\langle Q\rangle,\,P-\langle P\rangle\}\rangle\). Accordingly, \(\Delta Q\,\Delta P\) is a laboratory-frame diagnostic rather than a frame-invariant characterization of principal squeezing: a comparatively large value can reflect either genuinely enlarged uncertainty or a rotation of the noise ellipse away from the \(Q\)--\(P\) axes. Even so, the product remains a useful first indicator of how the laboratory-axis basins organize themselves in coefficient space.

\begin{figure}[t]
\centering
\includegraphics[width=\columnwidth]{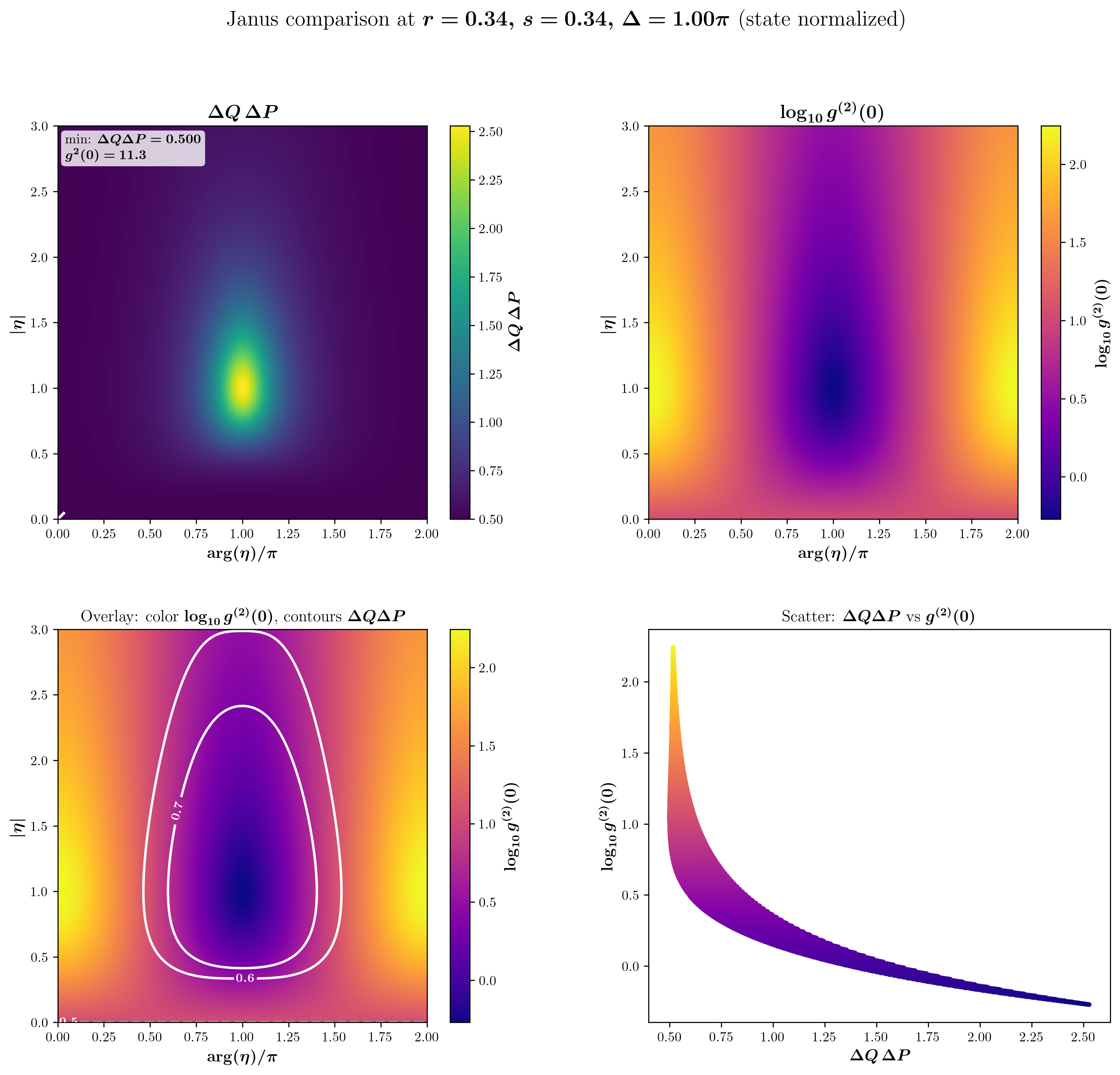}
\caption{Correlation comparison at $r=s=0.34$ and $\Delta=\pi$ (state normalized by Eq.~\eqref{eq:janus_norm_constraint}). The upper panels display the landscapes of $\Delta Q\,\Delta P$ and $\log_{10}g^{(2)}(0)$ over $(|\eta|,\delta)$. The lower-left overlay shows $\log_{10}g^{(2)}(0)$ with $\Delta Q\,\Delta P$ contours, while the lower-right panel shows the induced scatter relation $\log_{10}g^{(2)}(0)$ versus $\Delta Q\,\Delta P$.}
\label{fig:janus_compare_DQDP_vs_g2}
\end{figure}

Second moments do not, however, determine the photon statistics. To connect the near-Heisenberg basins to intensity fluctuations, it is useful to compare \(\Delta Q\,\Delta P\) with the equal-time correlation \(g^{(2)}(0)=\langle a^{\dagger 2}a^2\rangle/\langle a^\dagger a\rangle^2\). This comparison makes clear that a state may look nearly minimum-uncertainty in the fixed laboratory axes while still exhibiting strongly nontrivial photon statistics. At the same time, regions of very small mean occupation require some care in interpretation, since large values of \(g^{(2)}(0)\) can be amplified by the small denominator \(\langle a^\dagger a\rangle^2\). Figure~\ref{fig:janus_compare_DQDP_vs_g2} presents a representative case at \(\Delta=\pi\) and \(r=s=0.34\), showing the coefficient-space landscapes of \(\Delta Q\,\Delta P\) and \(\log_{10}g^{(2)}(0)\), an overlay of their contours, and the induced scatter relation between the two quantities. Taken together, these panels show that the approach to the near-Heisenberg basin need not coincide with suppressed intensity fluctuations; on the contrary, large bunching signatures can appear in nearby or overlapping regions of the same coefficient space.

\begin{figure*}[t]
\centering
\includegraphics[width=0.94\textwidth]{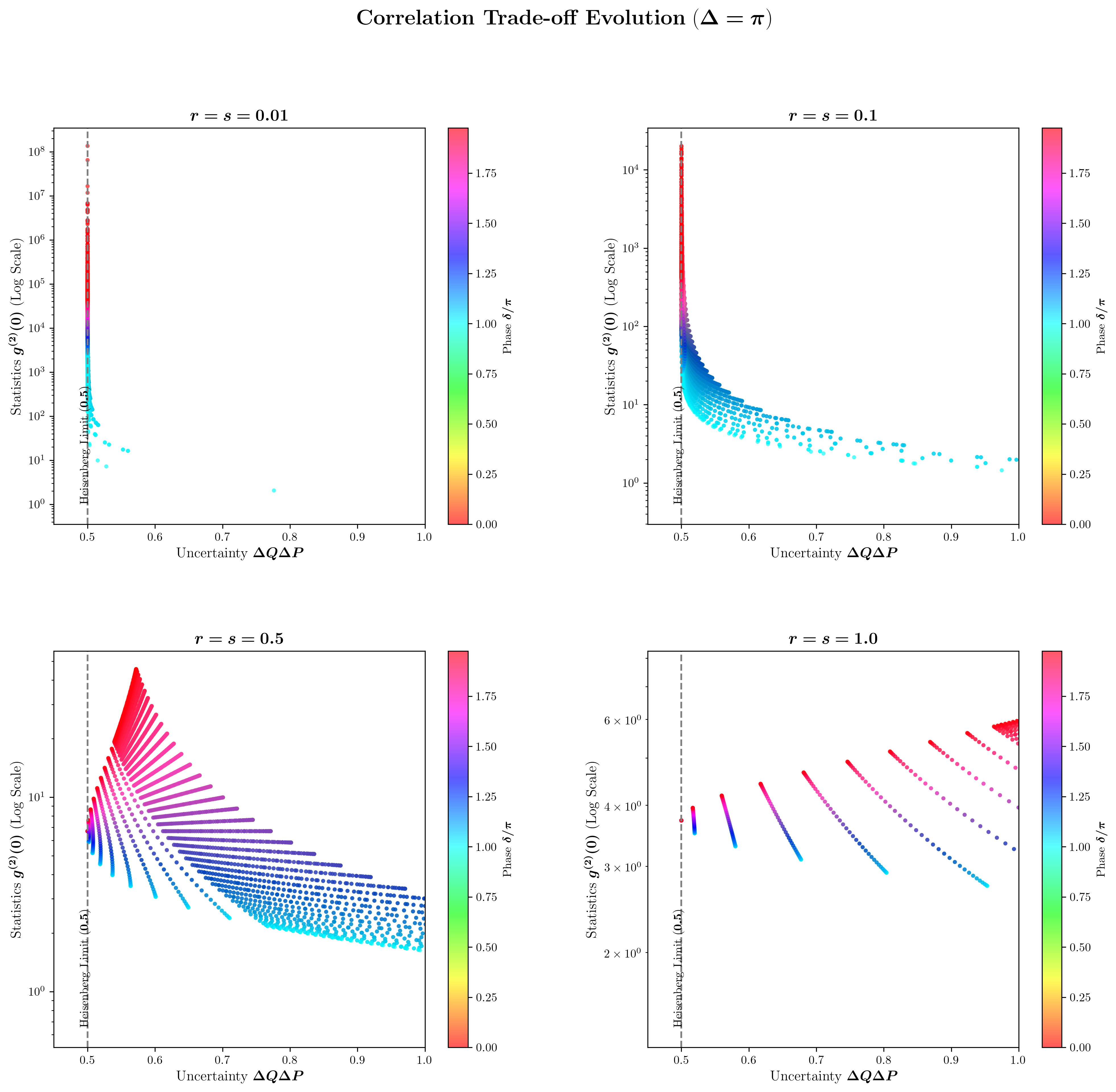}
\caption{Evolution of the correlation trade-off at $\Delta=\pi$: scatter plots of $\log_{10}g^{(2)}(0)$ versus $\Delta Q\,\Delta P$ for several choices of $r=s$. The vertical dashed line marks the Heisenberg boundary $\Delta Q\,\Delta P=1/2$, and the color encodes $\delta/\pi$.}
\label{fig:janus_tradeoff_evolution}
\end{figure*}

The same comparison may be repeated while varying the underlying squeezing strength at fixed \(\Delta=\pi\). In that way one can see not only how a single landscape is organized, but also how the entire correlation trade-off deforms as the constituents move deeper into the squeezed regime. Figure~\ref{fig:janus_tradeoff_evolution} therefore plots \(\log_{10}g^{(2)}(0)\) versus \(\Delta Q\,\Delta P\) for several equal-squeezing choices \(r=s\). The vertical line at \(\Delta Q\,\Delta P=1/2\) marks the minimum-uncertainty benchmark in the laboratory axes, while the color encodes the relative phase parameter \(\delta/\pi\). These plots emphasize that the coefficient-resolved relation between second-moment squeezing and photon statistics is itself strongly dependent on the squeezing scale.

It is also useful to examine how the coefficient-space basins themselves change as the squeezing strength is increased. At fixed \(\Delta=\pi\), the locations, widths, and curvatures of the low-\(\Delta Q\) and high-\(\Delta P\) regions evolve substantially with \(r=s\), indicating that the interference pattern is not merely deepened but reorganized.

\section{Quantum Fisher information and metrological bounds}
\label{sec:qfi_janus}

With the second-moment structure in hand, the remaining question is metrological: can Janus interference outperform the natural Gaussian benchmark once the comparison is made under a fair resource constraint? The relevant figure of merit is the quantum Fisher information, evaluated for a specified encoding model. We consider two settings. The first is the standard number-generated phase shift, for which the relevant quantities are the factorial moments \(N_1\) and \(N_2\). The second is quadratic-generator sensing, for which the sensitivity depends additionally on anomalous moments such as \(M_2\) and \(M_4\).

The point of this section is therefore not merely to show that Janus superpositions alter moments. The issue is whether those interference-induced changes translate into a larger QFI than the unique single squeezed vacuum under the same resource budget. We will consider two such budgets: fixed mean photon number \(\bar n\) and fixed homodyne-measured minimum quadrature variance \(V_{\min}\). The first is the standard energy benchmark, while the second is especially natural when squeezing is reported operationally in dB.

\subsection{\texorpdfstring{Unitary encodings and phase-shift sensing}{}}
\label{sec:qfi_definitions}

We consider unitary parameter encodings of the form \(\ket{\psi(\lambda)}=e^{-i\lambda G}\ket{\psi}\), where \(G\) is a parameter-independent Hermitian generator. For a pure probe state, the QFI reduces to

\begin{align}
F_Q(\lambda)=4\,\mathrm{Var}(G)
=4\Big(\bra{\psi}G^2\ket{\psi}-\big(\bra{\psi}G\ket{\psi}\big)^2\Big).
\label{eq:QFI_unitary_variance}
\end{align}

Thus, once the encoding generator is specified, metrological performance is determined entirely by the corresponding fluctuation structure of the probe.

A canonical single-mode example is a phase shift generated by photon number, \(\ket{\psi(\phi)}=e^{-i\phi \hat n}\ket{\psi}\), with \(\hat n\equiv a^\dagger a\). In that case, \(F_Q(\phi)=4\,\mathrm{Var}(\hat n)\), independent of \(\phi\). It is convenient to express the number variance in terms of factorial moments. Define

\begin{align}
N_k \equiv \bra{\psi}a^{\dagger k}a^k\ket{\psi}
=\big\langle \hat n(\hat n-1)\cdots(\hat n-k+1)\big\rangle,
\label{eq:factorial_moments_def}
\end{align}

together with the normalized equal-time intensity correlations
\begin{align*}
g^{(k)}(0)\equiv \frac{N_k}{N_1^{\,k}}.
\end{align*}
For \(k=1,2\), one finds

\begin{align}
\mathrm{Var}(\hat n)=N_2+N_1-N_1^2,
\qquad
F_Q(\phi)=4\big(N_2+N_1-N_1^2\big).
\label{eq:QFI_phase_factorial}
\end{align}

Equivalently, writing \(\bar n\equiv N_1\) and \(g^{(2)}(0)\equiv N_2/N_1^2\),

\begin{align}
F_Q(\phi)
=4\Big(\bar n+\big[g^{(2)}(0)-1\big]\bar n^2\Big).
\label{eq:QFI_phase_in_g2}
\end{align}

These equivalent forms will be the basic diagnostics for number-generated phase estimation below.

\subsection{\texorpdfstring{Phase estimation with Janus probes: factorial moments, fairness, and interference design rules}{}}
\label{sec:janus_qfi_and_g2}

For Janus probes, number-generated phase estimation is governed by the pair \((N_1,N_2)\). More generally, all factorial moments admit closed forms in terms of the squeezing polynomials \(P_k\) introduced in Ref.~\cite{Azizi2025Janus_higher}. For the normalized Janus state \(\ket{\psi}=\chi\ket{\xi}+\eta\ket{\zeta}\), one has

\begin{align}
N_k
&=
|\chi|^2\,\frac{P_k(x)}{(1-x)^k}
+
|\eta|^2\,\frac{P_k(y)}{(1-y)^k}
\label{eq:Nk_master_normalized}\\
&\quad
+2\,\mathrm{Re}\Bigg[
\chi\eta^\ast\,(1-x)^{1/4}(1-y)^{1/4}\,
\frac{P_k(z)}{(1-z)^{k+1/2}}
\Bigg],
\nn
\end{align}

where \(P_k\) is a degree-\(k\) polynomial with positive coefficients. The first two cases, \(P_1(u)=u\) and \(P_2(u)=2u^2+u\), give

\begin{align}
N_1
&=
|\chi|^2\frac{x}{1-x}
+|\eta|^2\frac{y}{1-y}
\label{eq:N1_janus}\\
&\quad
+2\,\mathrm{Re}\Bigg[
\chi\eta^\ast(1-x)^{1/4}(1-y)^{1/4}\,
\frac{z}{(1-z)^{3/2}}
\Bigg],
\nn
\end{align}

\begin{align}
N_2
&=
|\chi|^2\frac{x(2x+1)}{(1-x)^2}
+|\eta|^2\frac{y(2y+1)}{(1-y)^2}
\label{eq:N2_janus}\\
&\quad
+2\,\mathrm{Re}\Bigg[
\chi\eta^\ast(1-x)^{1/4}(1-y)^{1/4}\,
\frac{z(2z+1)}{(1-z)^{5/2}}
\Bigg].
\nn
\end{align}

The key structural point is that the interference term in \(N_2\) carries the more singular kernel \((1-z)^{-5/2}\), whereas the corresponding term in \(N_1\) carries \((1-z)^{-3/2}\). Janus interference can therefore modify number fluctuations more strongly than it modifies the mean photon number. Substituting Eqs.~\eqref{eq:N1_janus}--\eqref{eq:N2_janus} into Eq.~\eqref{eq:QFI_phase_factorial} gives a closed expression for the Janus phase-QFI.

The natural Gaussian reference is the single squeezed vacuum. In that limit, \(\bar n=\sinh^2 r\) and \(g^{(2)}_{\rm sq}(0)=3+1/\bar n\), so

\begin{align}
F_Q^{\rm sq}(\phi)
=
8\,\bar n(\bar n+1)
=
8\,\sinh^2 r\big(1+\sinh^2 r\big).
\label{eq:QFI_squeezed_vac}
\end{align}

Under a fixed-\(\bar n\) comparison, Janus beats this benchmark if and only if

\begin{align}
F_Q^{J}(\phi) > F_Q^{\rm sq}(\phi)
\quad\Longleftrightarrow\quad
g^{(2)}_{J}(0) > 3+\frac{1}{\bar n}.
\label{eq:Janus_beats_squeezed_condition}
\end{align}

Thus, at fixed mean photon number, improved phase sensitivity is equivalent to stronger bunching.

Without fixing \(\bar n\), however, \(g^{(2)}(0)\) by itself is not a fair metrological diagnostic, because interference can make it large by suppressing \(N_1\) while leaving \(N_2\) appreciable. In that case the meaningful quantity is the full QFI in Eq.~\eqref{eq:QFI_phase_factorial}, not \(g^{(2)}(0)\) alone.

The design principle behind this enhancement becomes especially transparent in the equal-strength family \(r=s\), for which \(x=y\) and \(z=xe^{i\Delta}\) with \(\Delta\neq 0\). Specializing Eqs.~\eqref{eq:N1_janus}--\eqref{eq:N2_janus} gives

\begin{align}
N_1
&=\frac{x}{1-x}\Bigg(
|\chi|^2+|\eta|^2
\nn\\
&\quad
+2\,\mathrm{Re}\Bigg[
\chi\eta^\ast e^{i\Delta}\,
\frac{(1-x)^{3/2}}{(1-xe^{i\Delta})^{3/2}}
\Bigg]
\Bigg),
\label{eq:N1_equal_strength_gap}
\end{align}

\begin{align}
N_2
&=\frac{x(2x+1)}{(1-x)^2}\Bigg(
|\chi|^2+|\eta|^2
\nn\\
&\quad
+2\,\mathrm{Re}\Bigg[
\chi\eta^\ast e^{i\Delta}\,
\frac{(1-x)^{5/2}}{(1-xe^{i\Delta})^{5/2}}\,
\frac{2xe^{i\Delta}+1}{2x+1}
\Bigg]
\Bigg).
\label{eq:N2_equal_strength_gap}
\end{align}

Defining the bracketed factors as \(\mathcal{A}_1\) and \(\mathcal{A}_2\), so that \(N_1=\frac{x}{1-x}\mathcal{A}_1\) and \(N_2=\frac{x(2x+1)}{(1-x)^2}\mathcal{A}_2\), one finds

\begin{align}
g^{(2)}_{J}(0)
=\frac{N_2}{N_1^2}
=\frac{2x+1}{x}\,
\frac{\mathcal{A}_2}{\mathcal{A}_1^2},
\label{eq:g2_equal_strength_gap}
\end{align}

and therefore, relative to the squeezed-vacuum baseline \(g^{(2)}_{\rm sq}(0)=(2x+1)/x\),

\begin{align}
\Delta g^{(2)}
=\frac{2x+1}{x}\Bigg(\frac{\mathcal{A}_2}{\mathcal{A}_1^2}-1\Bigg).
\label{eq:gap_in_terms_of_A_gap}
\end{align}

At fixed \(x\), enhancement is therefore driven by making \(\mathcal{A}_1\) small while preventing \(\mathcal{A}_2\) from vanishing.

This criterion also has a simple Fock-space interpretation. In the equal-strength family, the \(\ket{2}\) amplitude in \(\ket{\psi}\) is proportional to \(\chi e^{i\theta}+\eta e^{i\phi}=e^{i\phi}\big(\chi e^{i\Delta}+\eta\big)\). Hence exact cancellation of the two-photon sector occurs when

\begin{align}
\chi e^{i\Delta}+\eta = 0
\qquad\Longleftrightarrow\qquad
\frac{\eta}{\chi}=-e^{i\Delta}.
\label{eq:two_photon_cancellation_condition}
\end{align}

For \(\Delta\neq 0\ (\mathrm{mod}\ 2\pi)\), this cancellation does not remove the \(\ket{4}\) component. The operating point

\begin{align}
\Delta=\pi,
\qquad
\eta=\chi,
\label{eq:pi_phase_equal_weights}
\end{align}

therefore suppresses the \(\ket{2}\) sector while allowing the \(\ket{4}\) contribution to survive.

A normalized realization of this mechanism is obtained by taking \(\theta=0\), \(\phi=\pi\), so that \(z=-x\), and choosing \(\chi=\eta\) real. The normalization condition then gives

\begin{align}
\chi=\eta
=\frac{1}{\sqrt{2\Big(1+\sqrt{(1-x)/(1+x)}\Big)}}.
\label{eq:normalized_sum_pi_gap}
\end{align}

For this normalized family, the cancellation of the \(\ket{2}\) sector removes the linear contribution in \(x\), so both \(N_1\) and \(N_2\) begin at order \(x^2\). A direct expansion gives

\begin{align}
N_1
&=\frac{3}{2}x^2+\frac{13}{8}x^4+O(x^6),
\nn\\
N_2
&=\frac{9}{2}x^2+\frac{109}{8}x^4+O(x^6).
\label{eq:N1N2_smallx_corrected}
\end{align}

Accordingly,

\begin{align}
g^{(2)}_{J}(0)
=&\frac{N_2}{N_1^2}
=\frac{2}{x^2}+\frac{31}{18}+O(x^2),
\nn\\
g^{(2)}_{\rm sq}(0)
=&\frac{2x+1}{x}
=\frac{1}{x}+2.
\label{eq:g2_smallx_corrected}
\end{align}

Hence

\begin{align}
\Delta g^{(2)}
=\frac{2}{x^2}-\frac{1}{x}-\frac{5}{18}+O(x^2)
\longrightarrow \infty
\qquad
(x\to 0).
\label{eq:gap_unbounded_gap}
\end{align}

This makes explicit that \(\Delta g^{(2)}\) has no finite global maximum unless a resource constraint is imposed.

\subsection{\texorpdfstring{Quadratic-parameter estimation: squeezing generators and moment-reduced QFI}{}}
\label{sec:qfi_quadratic_generators}

Number-generated phase sensing depends only on number fluctuations, but many experimentally relevant encodings are quadratic in the field operators. In that case the QFI depends on fourth-order structure that is invisible to \(\mathrm{Var}(\hat n)\). Consider a unitary encoding \(U(\lambda)=e^{-i\lambda G}\), and let \(\vartheta\) denote a controllable phase specifying the generator axis. The corresponding quadratic generators may be written as

\begin{align}
G_{r} &\equiv \frac{i}{2}\Big(e^{-i\vartheta}a^{2}-e^{i\vartheta}a^{\dagger 2}\Big),\nn\\
G_{\vartheta} &\equiv \frac{1}{2}\Big(e^{-i\vartheta}a^{2}+e^{i\vartheta}a^{\dagger 2}\Big),
\label{eq:quadratic_generators_def}
\end{align}

These two generators probe, respectively, sensitivity to changes in squeezing strength and sensitivity to changes in squeezing-axis orientation. Their variances reduce to the four moments

\begin{align}
M_2\equiv &\langle a^2\rangle,\qquad
M_4\equiv\langle a^4\rangle,\nn\\
N_1\equiv & \langle a^\dagger a\rangle,\qquad
N_2\equiv\langle a^{\dagger 2}a^2\rangle,
\end{align}

together with the commutator
\[
[a^2,a^{\dagger 2}]=4a^\dagger a+2.
\]
The explicit reductions are collected in Appendix~\ref{app:var_quadratic_generators}. Using Eq.~\eqref{eq:QFI_unitary_variance}, one obtains

\begin{align}
F_Q^{(\vartheta)}
=&
2\,\Re\Big(e^{-2i\vartheta}\big(M_4-M_2^2\big)\Big) \nn\\
&
+2\Big(N_2+2N_1+1-|M_2|^{2}\Big),
\label{eq:QFI_quadratic_angle_moments}
\\
F_Q^{(r)}
=&
-2\,\Re\Big(e^{-2i\vartheta}\big(M_4-M_2^2\big)\Big) \nn\\
&
+2\Big(N_2+2N_1+1-|M_2|^{2}\Big).
\label{eq:QFI_quadratic_strength_moments}
\end{align}

The dependence on the axis phase \(\vartheta\) is isolated entirely in the connected fourth-order structure \(M_4-M_2^2\), so optimizing over \(\vartheta\) is immediate:

\begin{align}
F_{Q,\max}^{(\mathrm{quad})}
\equiv& \max_{\vartheta}\,F_Q^{(\vartheta)}
\label{eq:QFI_quad_phase_optimized_general}\\
=&
2\Big(N_2+2N_1+1-|M_2|^2\Big)
+2\big|M_4-M_2^2\big|.
\nn
\end{align}

The same optimized envelope is obtained if one starts from \(F_Q^{(r)}\), since the sign flip in the connected fourth-order term is absorbed by a shift of the generator-axis phase. This expression makes the non-Gaussian lever arm explicit. Even if two probes share the same second moments, so that \(N_1\) and \(|M_2|\) are fixed, a Janus superposition can still alter \(N_2\) and the connected fourth-order term \(M_4-M_2^2\), thereby changing the optimal quadratic-generator sensitivity.

\subsection{\texorpdfstring{Fixed measured squeezing as an operational benchmark}{}}
\label{sec:fixed_squeezing_db_benchmark}

For quadratic-generator metrology, fixing the mean photon number is not always the most operational comparison. In homodyne experiments, the quantity that is directly measured and routinely reported is the minimum quadrature noise, usually quoted in dB relative to vacuum. It is therefore natural to ask a different question: if two probes exhibit the same measured squeezing, can Janus interference still provide a metrological advantage? This benchmark is operational rather than globally resource-fair: it fixes the directly observed minimum quadrature noise, but it does not fix the mean photon number or the full fluctuation budget. It therefore asks whether Janus interference can outperform the squeezed-vacuum reference among probes with the same measured squeezing level.

The relevant experimental quantity is the minimum homodyne variance

\begin{align}
V_{\min}\equiv&
\min_{\varphi}\,\mathrm{Var}\big(X_\varphi\big),
\qquad
X_\varphi \equiv \frac{1}{\sqrt{2}}\big(ae^{-i\varphi}+a^\dagger e^{i\varphi}\big),
\nn\\
V_{\rm vac}=&\frac12.
\label{eq:Vmin_def}
\end{align}

It is convenient to normalize this by the vacuum level,

\begin{align}
u\equiv \frac{V_{\min}}{V_{\rm vac}}=2V_{\min},
\qquad
S_{\rm dB}\equiv -10\log_{10}u.
\label{eq:u_db_def}
\end{align}

For undisplaced even-parity probes, \(V_{\min}=(\Delta X)^2_{\min}\), so Eq.~\eqref{eq:principal_variances_in_terms_of_nbar_m} gives

\begin{align}
u
=
2(\Delta X)^2_{\min}
=
1+2N_1-2|M_2|.
\label{eq:u_moment_form}
\end{align}

Thus, at fixed measured squeezing, the benchmark parameter \(u\) is determined by the same low-order moments that govern the Janus second-moment structure.

The optimization over the homodyne phase \(\varphi\) used to define \(u\) is distinct from the optimization over the quadratic-generator axis \(\vartheta\) entering \(F_{Q,\max}^{(\mathrm{quad})}\). The benchmark therefore compares probes at equal measured squeezing, not at equal generator alignment.

In the squeezing regime \(0<u\le 1\), the natural pure-Gaussian reference is the unique single-mode squeezed vacuum with that same value of \(u\), up to phase-space rotation. For quadratic-generator sensing, its optimized QFI is obtained by expressing the squeezed-vacuum result in terms of \(u=e^{-2r}\), which gives

\begin{align}
F_{Q,\max}^{\rm sq}(u)
=
1+\frac12\Bigg(u^2+\frac{1}{u^2}\Bigg).
\label{eq:QFI_quad_sq_benchmark}
\end{align}

Accordingly, a Janus probe is advantageous at fixed measured squeezing precisely when

\begin{align}
F_{Q,\max}^{(\mathrm{quad})}
>
1+\frac12\Bigg(u^2+\frac{1}{u^2}\Bigg).
\label{eq:Janus_beats_sq_quadratic_condition}
\end{align}

Using Eq.~\eqref{eq:QFI_quad_phase_optimized_general}, this criterion may be written explicitly as

\begin{align}
2\Big(N_2&+2N_1+1-|M_2|^2\Big)
 \nn\\
&+
2\big|M_4-M_2^2\big|
>
1+\frac12\Bigg(u^2+\frac{1}{u^2}\Bigg).  \label{eq:Janus_beats_sq_quadratic_condition_explicit}
\end{align}

This makes the logic of the benchmark transparent: the left-hand side contains the full Janus fourth-order structure, whereas the right-hand side is the best pure-Gaussian sensitivity compatible with the same measured squeezing.

The same fixed-\(u\) viewpoint can also be applied to number-generated phase estimation. For an ideal squeezed vacuum, \(u=e^{-2r}\), so Eq.~\eqref{eq:QFI_squeezed_vac} becomes

\begin{align}
F_Q^{\rm sq}(u)
=
\frac12\Bigg(u-\frac{1}{u}\Bigg)^2.
\label{eq:QFI_sq_fixed_u}
\end{align}

For a Janus probe, the number-generated phase QFI is

\begin{align}
F_Q^{J}(\hat n)
=
4\big(N_2+N_1-N_1^2\big),
\label{eq:QFI_number_janus_fixed_u}
\end{align}

so the fixed-\(u\) benchmark is

\begin{align}
4\big(N_2+N_1-N_1^2\big)
>
\frac12\Bigg(u-\frac{1}{u}\Bigg)^2.
\label{eq:Janus_beats_sq_fixed_u}
\end{align}

Hence both the quadratic-generator problem and the number-generated phase problem can be compared against the same experimentally defined benchmark parameter, namely the observed squeezing level \(u\).

To make these benchmarks fully explicit for Janus states, the needed factorial moments are already given in Eqs.~\eqref{eq:N1_janus} and \eqref{eq:N2_janus}. The anomalous second moment is, from Eq.~\eqref{eq:janus_a2},

\begin{align}
M_2
=
-\Bigg[
&
|\chi|^2\frac{\alpha}{1-x}
+
|\eta|^2\frac{\beta}{1-y}
\nn\\
&
+\chi\eta^\ast(1-x)^{1/4}(1-y)^{1/4}
\frac{\alpha}{(1-z)^{3/2}}
\nn\\
&
+\chi^\ast\eta(1-x)^{1/4}(1-y)^{1/4}
\frac{\beta}{(1-z^\ast)^{3/2}}
\Bigg],
\label{eq:M2_Janus_fixed_u}
\end{align}

while the fourth-order anomalous moment follows from Eqs.~\eqref{eq:M4_Janus_assembly}, \eqref{eq:app_cross_even_final}, and \eqref{eq:app_cross_even_final_swapped}:

\begin{align}
M_4
=
3\Bigg[
&
|\chi|^2\frac{\alpha^2}{(1-x)^2}
+
|\eta|^2\frac{\beta^2}{(1-y)^2}
\nn\\
&
+\chi\eta^\ast(1-x)^{1/4}(1-y)^{1/4}
\frac{\alpha^2}{(1-z)^{5/2}}
\nn\\
&
+\chi^\ast\eta(1-x)^{1/4}(1-y)^{1/4}
\frac{\beta^2}{(1-z^\ast)^{5/2}}
\Bigg].
\label{eq:M4_Janus_fixed_u}
\end{align}

Equations~\eqref{eq:N1_janus}, \eqref{eq:N2_janus}, \eqref{eq:M2_Janus_fixed_u}, and \eqref{eq:M4_Janus_fixed_u} therefore provide a completely explicit analytic evaluation of both fixed-\(u\) benchmarks.

\begin{figure}[htbp]
\centering
\includegraphics[width=\columnwidth]{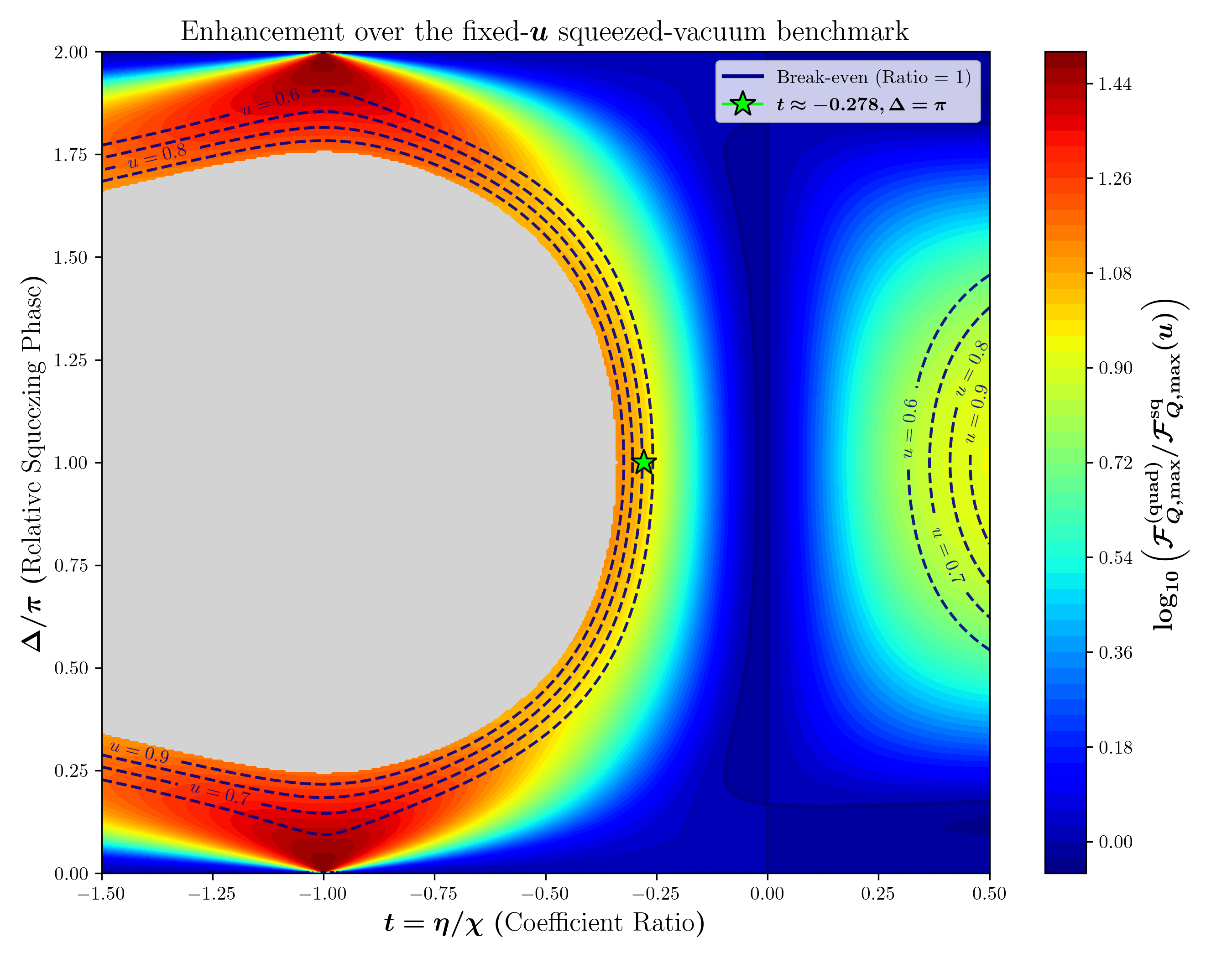}
\caption{Enhancement of the quadratic-generator quantum Fisher information for a Janus probe relative to the pure-Gaussian squeezed-vacuum benchmark evaluated at the same measured squeezing level \(u\). The landscape is plotted over the non-degenerate coefficient ratio \(t=\eta/\chi\) and the relative squeezing phase \(\Delta=\theta-\phi\) for fixed squeezing strengths \(x=y=1/2\). The gray region masks the non-squeezed regime (\(u > 1\)) where the squeezed-vacuum benchmark is not applicable. The solid contour identifies the break-even boundary where the ratio is 1 (logarithm is 0), while the dashed contours track constant values of the measured squeezing \(u\). The star indicates the representative state evaluated in Eq.~\eqref{eq:quad_advantage_example}.}
\label{fig:QFI_enhancement_fixed_u}
\end{figure}

The physical conclusion is then clear. At fixed measured squeezing, the Gaussian reference is completely determined by \(u\), whereas a Janus probe with the same \(u\) can still have substantially different values of \(N_2\) and of the connected fourth-order quantity \(M_4-M_2^2\). In the phase-opposed, nearly equal-strength regime \(r\simeq s\), \(\Delta=\theta-\phi\simeq\pi\), with a subtractive relative coefficient, interference can keep \(u\) moderate while strongly amplifying the fourth-order structure entering Eq.~\eqref{eq:QFI_quad_phase_optimized_general}. This is precisely the regime in which Janus probes can outperform the unique squeezed-vacuum benchmark even though the measured squeezing itself is held fixed. This operational advantage over the coefficient landscape is visualized in Fig.~\ref{fig:QFI_enhancement_fixed_u}.

A representative point is obtained for \(x=y=\tfrac12\), \(\theta=0\), \(\phi=\pi\), and normalized coefficients \(\chi\simeq 1.15\), \(\eta\simeq -0.32\) (marked by the star in Fig.~\ref{fig:QFI_enhancement_fixed_u}). Substituting these values into the explicit formulas above gives

\begin{align}
u\simeq 0.684,
\quad
F_{Q,\max}^{(\mathrm{quad})}\simeq 2.43\times 10^{1},
\quad
F_{Q,\max}^{\rm sq}(u)\simeq 2.30.
\label{eq:quad_advantage_example}
\end{align}

Thus the Janus probe exceeds the optimized single-squeezed-vacuum sensitivity by more than an order of magnitude at the same measured squeezing. This should be interpreted as an operational fixed-squeezing advantage rather than as a globally resource-fair improvement, because fixing \(u\) does not constrain the mean photon number. The fixed-\(u\) benchmark nevertheless shows that Janus interference can yield a substantial metrological gain over the Gaussian squeezed-vacuum reference even when the observed level of squeezing is held fixed.

\section{\texorpdfstring{Benchmark dependence: principal-axis no-go and constituent-relative compatibility}{}}
\label{sec:benchmark_dependence_no_go_vs_constituent}

The preceding subsections established metrological benchmarks for Janus probes under explicit resource constraints. Two logically distinct questions nevertheless remain. The first is a fair fixed-resource question: at fixed mean photon number, can a Janus state outperform the single squeezed vacuum in principal second-moment squeezing? The second is a span-constrained question: within a prescribed two-state squeezed-vacuum span, can a Janus superposition simultaneously outperform its two constituents in a fixed laboratory quadrature and in number-generated phase QFI? These questions are often conflated, but they are not equivalent. The first concerns the principal-axis quantity \((\Delta X)^2_{\min}\) under a global fixed-\(\bar n\) benchmark, whereas the second concerns a fixed laboratory-axis quantity such as \((\Delta Q)^2\) inside a prescribed span. The first has a negative answer, whereas the second has a positive one. The purpose of the present section is to make this distinction explicit and to show that the two conclusions are fully consistent. This resolution is summarized visually in Fig.~\ref{fig:Janus_no_contradiction_multipanel}.

\subsection{\texorpdfstring{A fixed-$\bar n$ no-go for principal second-moment squeezing}{}}
\label{sec:fixed_n_no_go_second_moments}

The interference-assisted reduction of \((\Delta Q)^2\) derived in Sec.~\ref{sec:janus_min_quadrature_variance} was a span-constrained statement: the two constituent squeezed vacua were held fixed, and only the superposition coefficients were varied. That optimization is therefore different from the fair metrological benchmark used in Sec.~\ref{sec:janus_qfi_and_g2}, where probes are compared at fixed mean photon number. A natural question is whether, under that fixed-\(\bar n\) benchmark, a Janus state can also exhibit a smaller principal quadrature variance than the single-mode squeezed vacuum with the same \(\bar n\). The answer is negative. At the level of second moments, the squeezed vacuum is extremal, and a genuinely non-Gaussian Janus superposition cannot do better.

For any undisplaced single-mode state, and in particular for the undisplaced Janus family, the moments introduced in Eq.~\eqref{eq:nbar_m_def} reduce to
\[
\bar n=\langle a^\dagger a\rangle,
\qquad
m=\langle a^2\rangle.
\]
The covariance matrix therefore takes the form given in Eq.~\eqref{eq:covariance_matrix_Janus},

\begin{align}
\mathbf{V}=
\begin{pmatrix}
\frac12+\bar n+\Re m & \Im m\\
\Im m & \frac12+\bar n-\Re m
\end{pmatrix}.
\end{align}

Its smaller eigenvalue is the principal minimum variance from Eq.~\eqref{eq:principal_variances_in_terms_of_nbar_m},

\begin{align}
(\Delta X)^2_{\min,J}
=
\frac12+\bar n-|m|.
\label{eq:DXmin_Janus_fixedn_subsec}
\end{align}

Hence the problem reduces entirely to determining how large \(|m|\) can be at fixed \(\bar n\).

The required bound follows directly from the positivity of quantum fluctuations. Define the centered quadratures

\begin{align}
\delta Q\equiv Q-\langle Q\rangle,
\qquad
\delta P\equiv P-\langle P\rangle,
\end{align}

and, for arbitrary complex numbers \(u\) and \(v\), consider the operator

\begin{align}
A=u\,\delta Q+v\,\delta P.
\end{align}

Because \(A^\dagger A\) is a positive operator, every physical state obeys

\begin{align}
\langle A^\dagger A\rangle\ge 0.
\label{eq:Apos_subsec}
\end{align}

Using \([Q,P]=i\), together with the decomposition of the mixed products into symmetric and antisymmetric parts,

\begin{align}
\langle \delta Q\,\delta P\rangle
&=
\frac12\langle\{\delta Q,\delta P\}\rangle
+
\frac12\langle[\delta Q,\delta P]\rangle
\nn\\
&=
V_{QP}+\frac{i}{2},
\\
\langle \delta P\,\delta Q\rangle
&=
\frac12\langle\{\delta Q,\delta P\}\rangle
-
\frac12\langle[\delta Q,\delta P]\rangle
\nn\\
&=
V_{QP}-\frac{i}{2},
\end{align}

one may rewrite Eq.~\eqref{eq:Apos_subsec} as

\begin{align}
\langle A^\dagger A\rangle
=
\begin{pmatrix}
u^\ast & v^\ast
\end{pmatrix}
\Bigg(
\mathbf{V}+\frac{i}{2}\Omega
\Bigg)
\begin{pmatrix}
u\\
v
\end{pmatrix}
\ge 0,
\label{eq:VplusOmega_positive_subsec}
\end{align}

where

\begin{align}
\Omega=
\begin{pmatrix}
0 & 1\\
-1 & 0
\end{pmatrix}.
\end{align}

Since Eq.~\eqref{eq:VplusOmega_positive_subsec} holds for all \(u\) and \(v\), the Hermitian matrix \(\mathbf{V}+\tfrac{i}{2}\Omega\) is positive semidefinite. For a \(2\times2\) Hermitian matrix, positive semidefiniteness implies a nonnegative determinant. Therefore

\begin{align}
0
\le
\det\Bigg(
\mathbf{V}+\frac{i}{2}\Omega
\Bigg)
&=
V_{QQ}V_{PP}-\Bigg(V_{QP}^2+\frac14\Bigg)
\nn\\
&=
\det\mathbf{V}-\frac14,
\end{align}

and hence

\begin{align}
\det\mathbf{V}\ge \frac14.
\label{eq:detV_bound_subsec}
\end{align}

This is the Robertson--Schr\"odinger uncertainty relation in covariance-matrix form.

Substituting the covariance matrix into Eq.~\eqref{eq:detV_bound_subsec} gives

\begin{align}
\det\mathbf{V}
&=
\Bigg(\frac12+\bar n+\Re m\Bigg)
\Bigg(\frac12+\bar n-\Re m\Bigg)
-(\Im m)^2
\nn\\
&=
\Bigg(\frac12+\bar n\Bigg)^2-|m|^2
\ge
\frac14.
\end{align}

Therefore

\begin{align}
|m|^2\le \bar n(\bar n+1),
\qquad
|m|\le \sqrt{\bar n(\bar n+1)}.
\label{eq:m_bound_fixedn_subsec}
\end{align}

Combining Eq.~\eqref{eq:m_bound_fixedn_subsec} with Eq.~\eqref{eq:DXmin_Janus_fixedn_subsec}, one obtains the universal lower bound

\begin{align}
(\Delta X)^2_{\min,J}
\ge
\frac12+\bar n-\sqrt{\bar n(\bar n+1)}.
\label{eq:Janus_principal_bound_fixedn_subsec}
\end{align}

The right-hand side is exactly the principal squeezed-quadrature variance of a single-mode squeezed vacuum at the same mean photon number. Indeed, for a squeezed vacuum, \(\bar n=\sinh^2 r\), and the anomalous moment has magnitude \(|m|=\sinh r\cosh r=\sqrt{\bar n(\bar n+1)}\), so Eq.~\eqref{eq:principal_variances_in_terms_of_nbar_m} yields

\begin{align}
(\Delta X)^2_{\min,\rm sq}
&=
\frac12+\bar n-\sqrt{\bar n(\bar n+1)}
\nn\\
&=
\frac12 e^{-2r}.
\label{eq:DXmin_sq_fixedn_subsec}
\end{align}

Hence

\begin{align}
(\Delta X)^2_{\min,J}\ge (\Delta X)^2_{\min,\rm sq}.
\label{eq:no_go_variance_fixedn_subsec}
\end{align}

The inequality is saturated only when \(|m|=\sqrt{\bar n(\bar n+1)}\), equivalently when \(\det\mathbf{V}=1/4\). For a single bosonic mode, saturation of the Robertson--Schr\"odinger bound corresponds to a minimum-uncertainty Gaussian state, namely a displaced squeezed state. In the present undisplaced even-parity setting, this reduces to the ordinary squeezed vacuum up to phase-space rotation. Hence a genuinely non-Gaussian Janus superposition cannot saturate Eq.~\eqref{eq:no_go_variance_fixedn_subsec}; strict equality occurs only in the squeezed-vacuum limit rather than for a genuine Janus superposition.

\begin{figure*}[htbp]
\centering
\includegraphics[width=\textwidth]{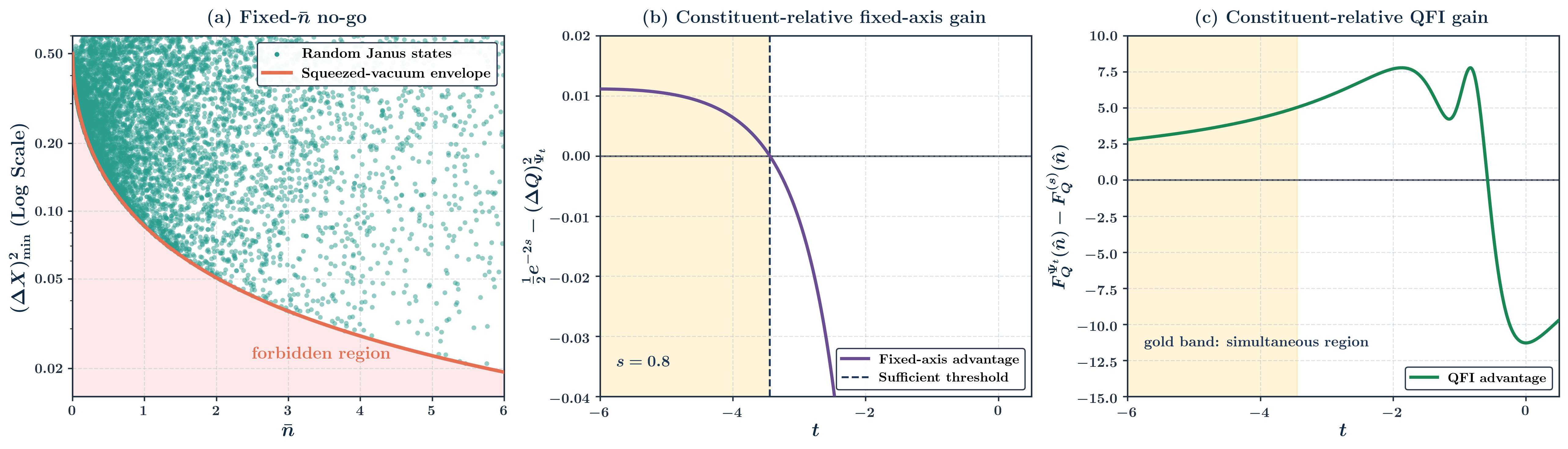}
\caption{Visual resolution of benchmark dependence for Janus interference. (a) Fixed-\(\bar{n}\) global no-go: the principal variance \((\Delta X)^2_{\min}\) of randomly generated Janus states is strictly bounded below by the pure squeezed-vacuum envelope, proving no global second-moment advantage exists at fixed energy. (b) and (c) Constituent-relative simultaneous advantage: using the auxiliary family \(|\Psi_t\rangle \propto |0\rangle + t|s\rangle\) with \(s=0.8\), the same subtractive superposition can simultaneously achieve a lower fixed-axis variance (panel b) and a higher number-generated phase QFI (panel c) than either constituent state. The gold band identifies the overlapping coefficient interval \(t\) where both local advantages coexist, demonstrating that constituent-relative gains do not contradict the global fixed-\(\bar{n}\) bound.}
\label{fig:Janus_no_contradiction_multipanel}
\end{figure*}

The physical content of this result is straightforward. At fixed mean photon number, principal second-moment squeezing is controlled entirely by how large \(|m|\) can become. The uncertainty principle restricts \(|m|\) through Eq.~\eqref{eq:m_bound_fixedn_subsec}, and the squeezed vacuum saturates that bound. Janus interference can redistribute noise within a prescribed two-state span and can even lower a fixed-axis variance below the better constituent value, as shown in Sec.~\ref{sec:janus_min_quadrature_variance}, but once the comparison is reformulated as a fixed-resource problem, the Gaussian squeezed vacuum remains extremal at the level of principal second moments. This global bound is illustrated in Fig.~\ref{fig:Janus_no_contradiction_multipanel}(a).

This also clarifies the relation to the number-generated phase-QFI results of Sec.~\ref{sec:janus_qfi_and_g2}. From Eq.~\eqref{eq:Janus_beats_squeezed_condition}, Janus exceeds the squeezed-vacuum benchmark for \(G=\hat n\) at fixed \(\bar n\) only when

\begin{align}
g_J^{(2)}(0)>3+\frac{1}{\bar n},
\end{align}

namely in a regime of stronger bunching. By contrast, Eq.~\eqref{eq:no_go_variance_fixedn_subsec} shows that no genuinely non-Gaussian Janus state can beat the single-mode squeezed vacuum in principal quadrature squeezing at the same \(\bar n\). Thus an enhancement of number-generated phase QFI does not arise from improved principal second-moment squeezing. The two advantages are distinct: the squeezed vacuum is optimal for principal quadrature variance, whereas Janus states can become advantageous for \(F_Q(\phi)\) only by reshaping higher-order fluctuation structure through interference.

This no-go is not the end of the story. It applies to a fixed-\(\bar n\) comparison against the globally optimal Gaussian reference and to the principal-axis variance \((\Delta X)^2_{\min}\). The next subsection changes both elements of the benchmark. There the comparison is constituent-relative rather than global, and the squeezing diagnostic is a fixed laboratory-axis variance rather than the principal minimum. In that narrower but still physically meaningful sense, a simultaneous Janus advantage does occur.

\subsection{\texorpdfstring{Constituent-relative compatibility of fixed-axis squeezing and number-generated phase QFI}{}}
\label{sec:simultaneous_constituent_beating}

With the fixed-\(\bar n\) no-go established, we now return to the span-constrained viewpoint of Sec.~\ref{sec:janus_min_quadrature_variance}. The question here is different. We no longer ask whether a Janus state can beat the single squeezed vacuum in principal second-moment squeezing under a fair global resource benchmark; the previous subsection excludes that possibility. Instead, we ask whether, within a prescribed two-state squeezed-vacuum span, one and the same Janus superposition can simultaneously beat its two constituents in a fixed laboratory quadrature and in number-generated phase QFI. The answer is affirmative.

Let
\begin{align}
\ket{\psi}=\chi\ket{\xi}+\eta\ket{\zeta},
\qquad
\braket{\psi}{\psi}=1.
\end{align}

Since both constituents are even-parity squeezed vacua, any superposition within their span is also even. Consequently \(\langle Q\rangle=0\) for the constituents and for the Janus superposition itself, so throughout this subsection the fixed-axis variance is simply the expectation value of \(Q^2\). The simultaneous constituent-beating conditions are therefore

\begin{align}
(\Delta Q)^2_J
&<
\min\Big\{
\bra{\xi}Q^2\ket{\xi},\,
\bra{\zeta}Q^2\ket{\zeta}
\Big\},
\label{eq:simultaneous_DQ_goal}
\\
F_Q^{J}(\hat n)
&>
\max\Big\{
F_Q^{\xi}(\hat n),\,
F_Q^{\zeta}(\hat n)
\Big\}.
\label{eq:simultaneous_QFI_goal}
\end{align}

For the individual squeezed-vacuum constituents, Eq.~\eqref{eq:QFI_squeezed_vac} gives

\begin{align}
F_Q^{\xi}(\hat n)
&=
8\sinh^2 r\big(1+\sinh^2 r\big),
\nn\\
F_Q^{\zeta}(\hat n)
&=
8\sinh^2 s\big(1+\sinh^2 s\big).
\label{eq:constituent_QFI_values}
\end{align}

The problem is thus to identify a normalized coefficient pair \((\chi,\eta)\) for which Eqs.~\eqref{eq:simultaneous_DQ_goal} and \eqref{eq:simultaneous_QFI_goal} hold simultaneously.

To analyze this compatibility in a setting where the mechanism is especially transparent, it is sufficient to restrict to the aligned family

\begin{align}
\theta=\phi=0,
\qquad
r\neq s,
\label{eq:aligned_family_condition}
\end{align}

for which both constituent squeezed vacua are squeezed along the same laboratory quadrature \(Q\). On this slice, the overlap variable introduced earlier reduces to

\begin{align}
z=\sqrt{xy}\in(0,1),
\end{align}

so the general formulas of Sec.~\ref{sec:janus_min_quadrature_variance} simplify immediately. In particular, Eq.~\eqref{eq:overlap_S_min} gives

\begin{align}
\mathcal S
=
(1-x)^{1/4}(1-y)^{1/4}(1-z)^{-1/2}
=
\frac{1}{\sqrt{\cosh(r-s)}},
\label{eq:S_aligned_family}
\end{align}

while the diagonal \(Q^2\) matrix elements reduce to

\begin{align}
A_Q
&=
\frac{1+x-2\sqrt{x}}{2(1-x)}
=
\frac12 e^{-2r},
\nn\\
B_Q
&=
\frac{1+y-2\sqrt{y}}{2(1-y)}
=
\frac12 e^{-2s}.
\label{eq:AQ_BQ_aligned}
\end{align}

For the off-diagonal element, Eq.~\eqref{eq:CQ_compact} gives

\begin{align}
C_Q
&=
(1-x)^{1/4}(1-y)^{1/4}
\frac{1+z-\sqrt{x}-\sqrt{y}}{2(1-z)^{3/2}}
\nn\\
&=
(1-x)^{1/4}(1-y)^{1/4}
\frac{(1-\sqrt{x})(1-\sqrt{y})}{2(1-z)^{3/2}}
\nn\\
&=
\mathcal S\,
\frac{(1-\sqrt{x})(1-\sqrt{y})}{2(1-z)}.
\label{eq:CQ_aligned_family}
\end{align}

Thus the generalized characteristic equation \(\det(\mathbf M_Q-\lambda\,\mathbf S)=0\) takes the form

\begin{align}
P(\lambda)
=
(A_Q-\lambda)(B_Q-\lambda)
-
\big(C_Q-\lambda \mathcal S\big)^2.
\label{eq:P_lambda_aligned}
\end{align}

Assume without loss of generality that \(r>s\), equivalently \(x>y\). Then

\begin{align}
A_Q<B_Q.
\label{eq:A_less_B_aligned}
\end{align}

Evaluating Eq.~\eqref{eq:P_lambda_aligned} at \(\lambda=A_Q\) gives

\begin{align}
P(A_Q)
&=
-\big(C_Q-A_Q\mathcal S\big)^2
\nn\\
&=
-\Bigg[
\mathcal S\,
\frac{(1-\sqrt{x})(\sqrt{x}-\sqrt{y})}
{2(1-z)(1+\sqrt{x})}
\Bigg]^2
<0.
\label{eq:P_of_A_negative}
\end{align}

Since \(1-\mathcal S^2>0\) for \(r\neq s\), the quadratic \(P(\lambda)\) opens upward, and \(P(\lambda)\to+\infty\) as \(\lambda\to-\infty\). Hence the smaller generalized eigenvalue satisfies

\begin{align}
\lambda_{Q,-}
<
A_Q
=
\min\{A_Q,B_Q\}.
\label{eq:strict_DQ_advantage_aligned}
\end{align}

Thus the aligned family already yields a strict constituent-beating reduction of the fixed-axis quadrature variance.

The corresponding minimizing generalized eigenvector may be chosen real, since \(A_Q\), \(B_Q\), \(C_Q\), and \(\mathcal S\) are all real on the aligned slice. Writing

\begin{align}
t_\ast\equiv \frac{\eta}{\chi},
\qquad
\chi\in\mathbb R_{>0},
\qquad
\eta\in\mathbb R,
\label{eq:tstar_def}
\end{align}

the first row of the generalized-eigenvalue equation gives

\begin{align}
t_\ast
=
-\,\frac{A_Q-\lambda_{Q,-}}
{C_Q-\lambda_{Q,-}\mathcal S}.
\label{eq:tstar_aligned}
\end{align}

The sign of \(t_\ast\) now follows directly. Since \(\lambda_{Q,-}<A_Q\), the numerator is positive. Moreover, Eq.~\eqref{eq:P_of_A_negative} implies

\begin{align}
C_Q-A_Q\mathcal S
=
\mathcal S\,
\frac{(1-\sqrt{x})(\sqrt{x}-\sqrt{y})}
{2(1-z)(1+\sqrt{x})}
>0.
\end{align}

Therefore

\begin{align}
C_Q-\lambda_{Q,-}\mathcal S
>
C_Q-A_Q\mathcal S
>0,
\end{align}

and hence

\begin{align}
t_\ast<0.
\label{eq:tstar_negative}
\end{align}

Accordingly, the fixed-axis variance is minimized by a subtractive superposition. The normalization condition then gives

\begin{align}
\chi_\ast
=
\frac{1}{\sqrt{1+t_\ast^2+2\mathcal S t_\ast}},
\qquad
\eta_\ast=t_\ast \chi_\ast.
\label{eq:chi_eta_star_aligned}
\end{align}

When \(\mathcal S\approx 1\), the two constituents are nearly linearly dependent, so the magnitudes of \(\chi_\ast\) and \(\eta_\ast\) may become large without any pathology; this simply reflects strong cancellation between two highly overlapping nonorthogonal states.

The QFI of this same coefficient-optimized state is obtained by specializing Eqs.~\eqref{eq:N1_janus} and \eqref{eq:N2_janus} to the same aligned slice. Since \(z=\sqrt{xy}\) is real and positive, all interference terms are real, and one finds

\begin{align}
N_1^\ast
=
\chi_\ast^2\Bigg[
\frac{x}{1-x}
+t_\ast^2\frac{y}{1-y}
+2t_\ast\,\mathcal S\,
\frac{\sqrt{xy}}{1-\sqrt{xy}}
\Bigg],
\label{eq:N1_star_aligned}
\end{align}

\begin{align}
N_2^\ast
=
\chi_\ast^2\Bigg[
&
\frac{x(2x+1)}{(1-x)^2}
+t_\ast^2\frac{y(2y+1)}{(1-y)^2}
\nn\\
&
+2t_\ast\,\mathcal S\,
\frac{\sqrt{xy}\big(2\sqrt{xy}+1\big)}
{\big(1-\sqrt{xy}\big)^2}
\Bigg],
\label{eq:N2_star_aligned}
\end{align}

and hence

\begin{align}
F_Q^{J,\ast}(\hat n)
=
4\Big(N_2^\ast+N_1^\ast-(N_1^\ast)^2\Big).
\label{eq:FQ_star_aligned}
\end{align}

The simultaneous constituent-beating question for the coefficient-optimized aligned state is therefore reduced to the pair of explicit inequalities

\begin{align}
\lambda_{Q,-}
<
\min\{A_Q,B_Q\},
\label{eq:simul_first_ineq_aligned}
\\
4\Big(N_2^\ast+N_1^\ast-(N_1^\ast)^2\Big)
>
\max\Big\{&
8\sinh^2 r(1+\sinh^2 r),   \nn\\
&
8\sinh^2 s(1+\sinh^2 s)
\Big\}.
\label{eq:simul_second_ineq_aligned}
\end{align}

The first inequality has already been established analytically by Eq.~\eqref{eq:strict_DQ_advantage_aligned}. To prove that the second can hold simultaneously, it suffices to exhibit one analytically controlled subfamily in which both constituent-beating conditions are satisfied at once.

\begin{figure*}[htbp]
\centering
\includegraphics[width=\textwidth]{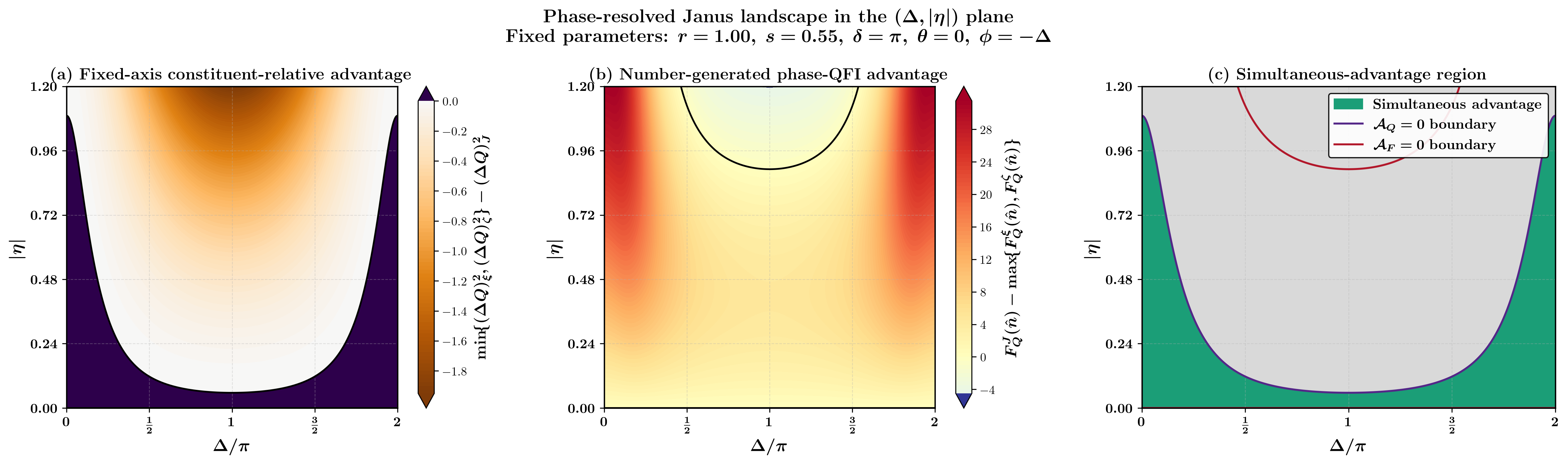}
\caption{Phase-resolved landscape demonstrating simultaneous constituent-relative advantage for the unequal-strength Janus family with \(r=1.00\) and \(s=0.55\). (a) Advantage in fixed-axis \(Q\)-squeezing compared to the better constituent limit. (b) Enhancement of the number-generated phase QFI over both constituent squeezed vacua. (c) The overlap region showing parameter space where the state is simultaneously quieter in the laboratory frame and more sensitive to phase shifts. The coefficient magnitude \(|\eta|\) and relative squeezing phase \(\Delta=\theta-\phi\) are scanned while maintaining subtractive superposition (\(\delta=\pi\)).}
\label{fig:Janus_phase_eta_landscape}
\end{figure*}

A particularly transparent subfamily is obtained by choosing one constituent to be the vacuum and the other to be a \(Q\)-squeezed vacuum with real squeezing parameter \(s>0\),

\begin{align}
\ket{0},
\qquad
\ket{s}\equiv S(s)\ket{0}.
\end{align}

Since the vacuum is itself a squeezed vacuum with zero squeezing, this remains within the Janus manifold. Consider the real one-parameter superposition

\begin{align}
\ket{\Psi_t}
=
\frac{\ket{0}+t\ket{s}}{\sqrt{\mathcal N(t)}},
\qquad
\mathcal N(t)\equiv 1+t^2+2\kappa t,
\label{eq:simul_aux_family}
\end{align}

where

\begin{align}
\kappa\equiv \braket{0}{s}=\frac{1}{\sqrt{\cosh s}}
\label{eq:simul_aux_overlap}
\end{align}

is the overlap from Eq.~\eqref{eq:sq_overlap_closed}. Since

\begin{align}
\mathcal N(t)=(t+\kappa)^2+(1-\kappa^2),
\end{align}

and \(0<\kappa<1\), the normalization denominator is strictly positive for all real \(t\). The relevant branch is again \(t<0\), corresponding to a subtractive superposition.

For this family,

\begin{align}
\bra{0}Q^2\ket{0}
&=
\frac12,
\qquad
\bra{s}Q^2\ket{s}=\frac12 e^{-2s},
\nn\\
\bra{0}Q^2\ket{s}
&=
\frac{\kappa}{1+e^{2s}},
\label{eq:simul_aux_Qmoments}
\end{align}

so that

\begin{align}
(\Delta Q)^2_{\Psi_t}
=
\frac{\frac12+t^2\frac12 e^{-2s}+2t\,\frac{\kappa}{1+e^{2s}}}
{\mathcal N(t)}.
\label{eq:simul_aux_DQ}
\end{align}

Subtracting the better constituent value \(\frac12 e^{-2s}\) gives

\begin{align}
(\Delta Q)^2_{\Psi_t}-\frac12 e^{-2s}
=
\frac{\frac12(1-e^{-2s})
+2t\Big(\frac{\kappa}{1+e^{2s}}-\frac{\kappa}{2e^{2s}}\Big)}
{\mathcal N(t)}.
\label{eq:simul_aux_DQdiff}
\end{align}

Since

\begin{align}
\frac{\kappa}{1+e^{2s}}-\frac{\kappa}{2e^{2s}}
=
\kappa\,
\frac{e^{2s}-1}{2e^{2s}(1+e^{2s})}
>0,
\end{align}

one obtains the sufficient condition

\begin{align}
(\Delta Q)^2_{\Psi_t}<\frac12 e^{-2s}
\qquad\text{whenever}\qquad
t<-\frac{1+e^{2s}}{2\kappa}.
\label{eq:simul_aux_DQcond}
\end{align}

Hence this auxiliary family contains a nonempty interval in which the Janus state beats both constituents in fixed-axis \(Q\)-noise.

The same family also makes the QFI mechanism transparent. Because \(\hat n\ket{0}=0\), one has

\begin{align}
\bra{0}\hat n\ket{s}=0,
\qquad
\bra{0}\hat n^2\ket{s}=0,
\label{eq:simul_aux_ncross}
\end{align}

so the moments reduce to

\begin{align}
\langle \hat n\rangle_{\Psi_t}
&=
\frac{t^2 n_s}{\mathcal N(t)},
\qquad
n_s=\sinh^2 s,
\\
\langle \hat n^2\rangle_{\Psi_t}
&=
\frac{t^2 \mu_{2,s}}{\mathcal N(t)},
\qquad
\mu_{2,s}=3n_s^2+2n_s.
\label{eq:simul_aux_nmoments}
\end{align}

Defining

\begin{align}
\Lambda(t)\equiv \frac{t^2}{\mathcal N(t)},
\end{align}

one finds

\begin{align}
F_Q^{\Psi_t}(\hat n)
=
4\Big[\Lambda(t)\,\mu_{2,s}-\Lambda(t)^2 n_s^2\Big].
\label{eq:simul_aux_FQ}
\end{align}

Subtracting the squeezed-constituent value \(F_Q^{(s)}(\hat n)=8n_s(n_s+1)\) yields

\begin{align}
F_Q^{\Psi_t}(\hat n)-F_Q^{(s)}(\hat n)
=
4\big(\Lambda-1\big)
\Big[\mu_{2,s}-n_s^2(\Lambda+1)\Big].
\label{eq:simul_aux_FQdiff}
\end{align}

As \(t\to-\infty\), one has \(\Lambda(t)\to 1^+\). Therefore, for sufficiently large negative \(t\),

\begin{align}
\Lambda-1>0,
\qquad
\mu_{2,s}-n_s^2(\Lambda+1)\to n_s^2+2n_s>0,
\end{align}

and hence

\begin{align}
F_Q^{\Psi_t}(\hat n)>F_Q^{(s)}(\hat n)>F_Q^{(0)}(\hat n)=0.
\label{eq:simul_aux_QFIcond}
\end{align}

Because Eq.~\eqref{eq:simul_aux_DQcond} and Eq.~\eqref{eq:simul_aux_QFIcond} both hold for sufficiently large negative \(t\), their overlap is nonempty. The auxiliary family therefore contains a finite interval of subtractive superpositions for which the same Janus state simultaneously beats both constituents in fixed-axis squeezing and in number-generated phase QFI.

The relation between the present result and the previous subsection is now clear. The no-go result of Sec.~\ref{sec:fixed_n_no_go_second_moments} is a fixed-\(\bar n\) statement about the principal quadrature variance \((\Delta X)^2_{\min}\), compared against the unique single squeezed vacuum with the same mean photon number. By contrast, the result established here is constituent-relative: the Janus state is compared only with the two squeezed vacua that generate its span, and the squeezing diagnostic is the fixed laboratory-axis variance \((\Delta Q)^2\), not the principal-axis minimum. These are different questions and different benchmarks, and there is therefore no contradiction between the two conclusions.

Taken together, the two subsections show that Janus interference does not surpass the Gaussian squeezed vacuum in principal second-moment squeezing under a fair fixed-\(\bar n\) comparison. It can nevertheless produce a simultaneous constituent-relative advantage in fixed-axis squeezing and number-generated phase QFI within a prescribed two-state span. The extent of this constituent-relative gain across the underlying coefficient space is further visualized in Fig.~\ref{fig:Janus_phase_eta_landscape}, showing the advantage survives when varying the underlying constituent phases. The metrological gain therefore does not arise from beating the globally optimal principal second-moment squeezing benchmark. Rather, it is a span-constrained interference effect: fixed-axis noise can be reduced relative to the chosen constituents, while the number-generated phase QFI is enhanced through the interference-sensitive higher-order fluctuation structure encoded in the factorial moments.

\subsection{\texorpdfstring{An exact normalized Janus family that beats the fixed-\(\bar n\) squeezed-vacuum phase-QFI benchmark}{}}
\label{sec:exact_fixed_nbar_beating_family}

Equation~\eqref{eq:Janus_beats_squeezed_condition} already reduces the fixed-\(\bar n\) phase-metrology comparison to a statement about \(g^{(2)}(0)\). It is therefore useful to exhibit an explicit normalized Janus subfamily for which the advantage can be established in closed form. A particularly simple choice is the vacuum--squeezed family
\begin{align}
\ket{\psi}
=
\chi\ket{0}
+
|\eta|e^{i\delta}\ket{s},
\qquad
\chi\in\mathbb R_{\ge 0},
\qquad
s>0,
\label{eq:exact_family_state}
\end{align}
with \(\ket{s}\equiv S(s)\ket{0}\). Since the state is normalized, Eq.~\eqref{eq:janus_norm_constraint} gives
\begin{align}
\chi^2+|\eta|^2+2\kappa\,\chi|\eta|\cos\delta=1,
\qquad
\kappa\equiv\braket{0}{s}=\frac{1}{\sqrt{\cosh s}},
\label{eq:exact_family_norm}
\end{align}
so that the nonnegative branch is
\begin{align}
\chi
=
-\kappa|\eta|\cos\delta
+
\sqrt{1-|\eta|^2+\kappa^2|\eta|^2\cos^2\delta}.
\label{eq:exact_family_chi_solution}
\end{align}
The radicand is strictly positive for every \(0<|\eta|<1\), so this normalized family exists on the full open interval \(0<|\eta|<1\) for any fixed coefficient phase \(\delta\).

For this subfamily, the general Janus invariants specialize to
\begin{align}
x=0,
\qquad
y=\tanh^2 s,
\qquad
z=0.
\label{eq:exact_family_xyz}
\end{align}
Equivalently,
\begin{align}
\bra{0}\hat n\ket{s}=0,
\qquad
\bra{0}\hat n(\hat n-1)\ket{s}=0,
\label{eq:exact_family_cross_zero}
\end{align}
so the vacuum--squeezed cross terms vanish identically in both \(N_1\) and \(N_2\). Using Eqs.~\eqref{eq:N1_janus} and \eqref{eq:N2_janus}, one finds
\begin{align}
N_1
=
|\eta|^2\sinh^2 s,
\qquad
N_2
=
|\eta|^2\sinh^2 s\big(3\sinh^2 s+1\big).
\label{eq:exact_family_N1N2}
\end{align}
Hence
\begin{align}
\bar n
=
N_1
=
|\eta|^2\sinh^2 s,
\label{eq:exact_family_nbar}
\end{align}
and therefore, at fixed \(\bar n\),
\begin{align}
\sinh^2 s=\frac{\bar n}{|\eta|^2}.
\label{eq:exact_family_s_fixed_nbar}
\end{align}

The corresponding equal-time intensity correlation is
\begin{align}
g_J^{(2)}(0)
=
\frac{N_2}{N_1^2}
=
\frac{1}{|\eta|^2}
\Bigg(
3+\frac{1}{\sinh^2 s}
\Bigg)
=
\frac{3}{|\eta|^2}+\frac{1}{\bar n}.
\label{eq:exact_family_g2}
\end{align}
At the same mean photon number, the squeezed-vacuum reference is
\begin{align}
g_{\rm sq}^{(2)}(0)
=
3+\frac{1}{\bar n}.
\label{eq:exact_family_g2_sq}
\end{align}
Thus
\begin{align}
g_J^{(2)}(0)-g_{\rm sq}^{(2)}(0)
=
3\Bigg(\frac{1}{|\eta|^2}-1\Bigg)
>0
\qquad
(0<|\eta|<1).
\label{eq:exact_family_g2_gap}
\end{align}
Using Eq.~\eqref{eq:QFI_phase_in_g2}, the phase QFI follows immediately:
\begin{align}
F_Q^{J}(\hat n)
&=
4\Big(\bar n+\big[g_J^{(2)}(0)-1\big]\bar n^2\Big)
\nn\\
&=
8\bar n(\bar n+1)
+
12\bar n^2\Bigg(\frac{1}{|\eta|^2}-1\Bigg).
\label{eq:exact_family_QFI_J}
\end{align}
Since the squeezed-vacuum benchmark at fixed \(\bar n\) is \(F_Q^{\rm sq}(\hat n)=8\bar n(\bar n+1)\), one obtains
\begin{align}
F_Q^{J}(\hat n)-F_Q^{\rm sq}(\hat n)
=
12\bar n^2\Bigg(\frac{1}{|\eta|^2}-1\Bigg)
>0
\qquad
(0<|\eta|<1).
\label{eq:exact_family_QFI_gap_fixed_nbar}
\end{align}
Therefore,
\begin{align}
F_Q^{J}(\hat n)>F_Q^{\rm sq}(\hat n)
\qquad\text{for every}\qquad
0<|\eta|<1.
\label{eq:exact_family_QFI_advantage}
\end{align}

This subfamily thus provides an exact existence proof of a normalized Janus advantage in number-generated phase estimation under the fair fixed-\(\bar n\) benchmark. The mechanism is transparent. Because the vacuum--squeezed cross terms vanish in \(N_1\) and \(N_2\), both factorial moments scale linearly with \(|\eta|^2\), whereas the normalized ratio \(g^{(2)}(0)=N_2/N_1^2\) acquires the inverse factor \(1/|\eta|^2\). At fixed \(\bar n\), this enhances the higher-order fluctuation term in Eq.~\eqref{eq:QFI_phase_in_g2} while leaving the squeezed-vacuum baseline \(8\bar n(\bar n+1)\) unchanged.

The role of the coefficient phase \(\delta\) is equally simple. Since the cross contributions to \(N_1\) and \(N_2\) vanish identically, the quantities \(g_J^{(2)}(0)\), \(F_Q^{J}(\hat n)\), and the gap in Eq.~\eqref{eq:exact_family_QFI_gap_fixed_nbar} are independent of \(\delta\). The phase enters only through the normalization in Eq.~\eqref{eq:exact_family_chi_solution}. In particular, for the \(\delta=0\) slice one has
\begin{align}
\chi
=
-\kappa|\eta|
+
\sqrt{1-(1-\kappa^2)|\eta|^2},
\label{eq:exact_family_chi_delta_zero}
\end{align}
whereas for arbitrary nonzero \(\delta\),
\begin{align}
\chi
=
-\kappa|\eta|\cos\delta
+
\sqrt{1-|\eta|^2+\kappa^2|\eta|^2\cos^2\delta}.
\label{eq:exact_family_chi_delta_general}
\end{align}
Hence the fixed-\(\bar n\) phase-QFI advantage persists for arbitrary coefficient phase, provided the state is normalized and \(0<|\eta|<1\).

There is, however, an important interpretive point. At fixed \(\bar n\), Eq.~\eqref{eq:exact_family_s_fixed_nbar} implies \(\sinh^2 s=\bar n/|\eta|^2\), so as \(|\eta|\to 0\) the squeezed constituent becomes arbitrarily strong while its coefficient becomes arbitrarily small. Accordingly, the gain in Eq.~\eqref{eq:exact_family_QFI_gap_fixed_nbar} grows without bound in that limit. This does not invalidate the fixed-\(\bar n\) comparison, but it shows that mean photon number alone does not upper-bound the phase QFI over such non-Gaussian superposition families. The present subsection should therefore be read as an exact existence proof rather than as a statement of optimality under a more restrictive resource model.

This result should also be read together with Sec.~\ref{sec:fixed_n_no_go_second_moments}. The family above yields a genuine fixed-\(\bar n\) Janus advantage for \(F_Q(\hat n)\), but not for principal second-moment squeezing. The squeezed vacuum remains extremal for \((\Delta X)^2_{\min}\) at fixed \(\bar n\), whereas the present advantage arises entirely from higher-order number fluctuations.

\section{Conclusion}
\label{sec:conclusion}

In this work, we analyzed the quadrature structure and metrological performance of Janus states, defined as coherent superpositions of two single-mode squeezed vacua with independently tunable squeezing strengths and phases. Because this family remains analytically tractable while already exhibiting genuinely non-Gaussian interference effects, it provides a useful setting in which second moments, factorial moments, and the low-order anomalous moments relevant for quantum Fisher information can all be evaluated in closed form.

At the level of quadrature fluctuations, we derived explicit formulas for the laboratory-axis variances, the covariance matrix, and the principal squeezing obtained by homodyne optimization. We also formulated the coefficient-optimized fixed-axis problem within a prescribed two-state span as a generalized eigenvalue problem, making it possible to identify the regime in which interference lowers a chosen laboratory quadrature below the better constituent value. This clarified the distinction between principal-axis squeezing of a fixed state and span-constrained fixed-axis squeezing produced by coherent superposition.

The metrological analysis showed that these interference effects must be interpreted relative to the benchmark being used. Under a fair fixed-mean-photon-number comparison, the single squeezed vacuum remains extremal for principal second-moment squeezing, yielding a no-go result for any genuine Janus advantage at that level. By contrast, within a prescribed two-state span, a Janus superposition can simultaneously beat its constituents in a fixed laboratory quadrature and in number-generated phase QFI. We also showed that under the operational benchmark of fixed measured squeezing, Janus interference can substantially enhance quadratic-generator QFI relative to the pure-Gaussian squeezed-vacuum reference with the same observed squeezing level.

Taken together, these results show that quadrature squeezing, higher-order correlations, and metrological sensitivity are related but distinct notions, and that their ordering depends crucially on the comparison being made. Janus superpositions therefore provide a controlled non-Gaussian platform for studying benchmark-dependent quantum advantage and for exploring interference-based enhancements beyond the Gaussian squeezed-vacuum setting.


\section*{Acknowledgments}
I am grateful to Girish Agarwal and Suhail Zubairy for discussions. This work was supported by the Robert A. Welch Foundation (Grant No. A-1261).


\appendix

\onecolumngrid


\section{\texorpdfstring{Closed form for $\bra{\zeta}a^{2k}\ket{\xi}$}{}}
\label{app:cross_even_moments}

This appendix derives a closed expression for the even-order cross moment between two squeezed vacua, \(\bra{\zeta}a^{2k}\ket{\xi}\) with \(k\in\mathbb{N}_0\), in the same invariant notation used throughout the main text.

\begin{align}
\xi=r e^{i\theta},\qquad
\zeta=s e^{i\phi},\qquad
\alpha\equiv\tanh r\,e^{i\theta},\qquad
\beta\equiv\tanh s\,e^{i\phi},\qquad
x\equiv|\alpha|^2,\qquad
y\equiv|\beta|^2,\qquad
z\equiv \alpha\beta^\ast,
\quad
\Delta\equiv\theta-\phi.
\label{eq:app_xyz_def}
\end{align}

In the disentangled representation,

\begin{align}
\ket{\xi}
&=(1-x)^{1/4}\exp\Big(-\frac{\alpha}{2}a^{\dagger 2}\Big)\ket{0},\nn\\
\bra{\zeta}
&=(1-y)^{1/4}\bra{0}\exp\Big(-\frac{\beta^\ast}{2}a^{2}\Big).
\label{eq:app_sq_disentangled}
\end{align}

The overlap \(\braket{\zeta}{\xi}=(1-x)^{1/4}(1-y)^{1/4}(1-z)^{-1/2}\) is recovered as the \(k=0\) case of the general result below.

\subsection{\texorpdfstring{Fock expansions and reduction to a single series}{}}
\label{app:cross_even_series}

Using the standard even-Fock expansion of a squeezed vacuum,

\begin{align}
\ket{\xi}
=(1-x)^{1/4}\sum_{n=0}^{\infty}
\Big(-\alpha\Big)^{n}\,
\frac{(2n-1)!!}{\sqrt{(2n)!}}\,
\ket{2n},
\label{eq:app_sq_fock_ket}
\end{align}

and its bra counterpart,

\begin{align}
\bra{\zeta}
=(1-y)^{1/4}\sum_{m=0}^{\infty}
\Big(-\beta^\ast\Big)^{m}\,
\frac{(2m-1)!!}{\sqrt{(2m)!}}\,
\bra{2m},
\label{eq:app_sq_fock_bra}
\end{align}

we use the standard conventions \((0)!!=1\) and \((-1)!!=1\), so that the \(n=0\) or \(m=0\) term equals unity. Together with

\begin{align}
a^{2k}\ket{2n}=\sqrt{\frac{(2n)!}{(2n-2k)!}}\ket{2n-2k},
\qquad n\ge k,
\label{eq:app_a2k_on_fock}
\end{align}

one finds

\begin{align}
\bra{\zeta}a^{2k}\ket{\xi}
=&\,
(1-x)^{1/4}(1-y)^{1/4}
\sum_{m=0}^{\infty}\sum_{n=k}^{\infty}
\Big(-\beta^\ast\Big)^{m}
\Big(-\alpha\Big)^{n}\,
\frac{(2m-1)!!}{\sqrt{(2m)!}}\,
\frac{(2n-1)!!}{\sqrt{(2n-2k)!}}
\,\braket{2m}{2n-2k}.
\label{eq:app_double_sum_start}
\end{align}

The inner product enforces \(m=n-k\), and the double sum collapses to a single series,

\begin{align}
\bra{\zeta}a^{2k}\ket{\xi}
=&\,
(1-x)^{1/4}(1-y)^{1/4}
\Big(-\alpha\Big)^{k}
\sum_{m=0}^{\infty}
z^{m}\,
\frac{(2m+2k-1)!!}{(2m)!!},
\label{eq:app_reduced_series}
\end{align}

where \(z\) is defined in Eq.~\eqref{eq:app_xyz_def} and we used \((2m)!=(2m)!!(2m-1)!!\).

\subsection{\texorpdfstring{Generating function and closed form}{}}
\label{app:cross_even_closed_form}

Introduce the generating series

\begin{align}
G_{2k}(z)\equiv \sum_{m=0}^{\infty} z^{m}\,\frac{(2m+2k-1)!!}{(2m)!!}.
\label{eq:app_G_def}
\end{align}

The series \(G_{2k}(z)\) converges for \(|z|<1\), which holds automatically here since \(|z|=\tanh r\,\tanh s<1\) for finite squeezing. For \(k=0\) with the convention \((-1)!!=1\), this reduces to the standard binomial series

\begin{align}
G_{0}(z)=\sum_{m=0}^{\infty} z^{m}\,\frac{(2m-1)!!}{(2m)!!}=(1-z)^{-1/2}.
\label{eq:app_G0}
\end{align}

The higher-\(k\) series can be generated by repeatedly pulling down the factor \((2m+2k-1)\) using the identity \(2m\,z^{m}=(2z\partial_{z})z^{m}\). This yields the operator form

\begin{align}
G_{2k}(z)
=\Big(2z\frac{\partial}{\partial z}+2k-1\Big)\cdots
\Big(2z\frac{\partial}{\partial z}+1\Big)\,G_{0}(z),
\label{eq:app_G_operator}
\end{align}

and, equivalently, the recursion

\begin{align}
G_{2k+2}(z)=\Big(2z\frac{\partial}{\partial z}+2k+1\Big)\,G_{2k}(z).
\label{eq:app_G_recursion}
\end{align}

Applying this recursion to \(G_0(z)=(1-z)^{-1/2}\) gives the closed form

\begin{align}
G_{2k}(z)=\frac{(2k-1)!!}{(1-z)^{k+1/2}},
\label{eq:app_G_closed}
\end{align}

where the fractional power is taken on the principal branch for complex \(z\).

A direct inductive verification follows from Eq.~\eqref{eq:app_G_recursion}. The base case \(k=0\) is Eq.~\eqref{eq:app_G0}. Assuming Eq.~\eqref{eq:app_G_closed} holds for some \(k\ge 0\), one has

\begin{align}
G_{2k+2}(z)
&=\Big(2z\frac{\partial}{\partial z}+2k+1\Big)G_{2k}(z)\nn\\
&=\Big(2z\frac{\partial}{\partial z}+2k+1\Big)\Big[(2k-1)!!(1-z)^{-k-1/2}\Big]\nn\\
&=(2k-1)!!(2k+1)(1-z)^{-k-3/2}\nn\\
&=(2k+1)!!\,(1-z)^{-(k+1)-1/2},
\end{align}

which is Eq.~\eqref{eq:app_G_closed} with \(k\mapsto k+1\).

Substituting Eq.~\eqref{eq:app_G_closed} into Eq.~\eqref{eq:app_reduced_series} gives the final cross moment,

\begin{align}
\bra{\zeta}a^{2k}\ket{\xi}
=(1-x)^{1/4}(1-y)^{1/4}\,(2k-1)!!\,
\frac{\Big(-e^{i\theta}\tanh r\Big)^{k}}{(1-z)^{k+1/2}},
\qquad k=0,1,2,\ldots
\label{eq:app_cross_even_final}
\end{align}

For consistency, \(k=0\) reproduces the overlap, while \(k=1,2\) reproduce the special cases \(\bra{\zeta}a^{2}\ket{\xi}\propto(1-z)^{-3/2}\) and \(\bra{\zeta}a^{4}\ket{\xi}\propto(1-z)^{-5/2}\) used in the main text.

The swapped matrix element follows by exchanging \((x,\theta)\leftrightarrow(y,\phi)\) and \(z\leftrightarrow z^\ast\),

\begin{align}
\bra{\xi}a^{2k}\ket{\zeta}
=(1-x)^{1/4}(1-y)^{1/4}\,(2k-1)!!\,
\frac{\Big(-e^{i\phi} \tanh s\Big)^{k}}{(1-z^\ast)^{k+1/2}}.
\label{eq:app_cross_even_final_swapped}
\end{align}


\section{\texorpdfstring{Variances of the quadratic generators $G_r$ and $G_{\vartheta}$}{}}
\label{app:var_quadratic_generators}

This appendix records the variance reductions used in Sec.~\ref{sec:qfi_quadratic_generators}. Expectation values are taken in an arbitrary probe state \(\ket{\psi}\), with \(\langle O\rangle\equiv\bra{\psi}O\ket{\psi}\). The moments used throughout are

\begin{align}
M_2\equiv \langle a^2\rangle,\qquad
M_4\equiv \langle a^4\rangle,\qquad
N_1\equiv \langle a^\dagger a\rangle,\qquad
N_2\equiv \langle a^{\dagger 2}a^2\rangle.
\label{eq:app_MN_defs}
\end{align}

\subsection{\texorpdfstring{Variance of $G_r$}{}}

With the generator definition

\begin{align}
G_{r}=\frac{i}{2}\Big(e^{-i\vartheta}a^{2}-e^{i\vartheta}a^{\dagger 2}\Big),
\label{eq:app_Gr_def}
\end{align}

the variance is \(\mathrm{Var}(G_r)=\langle G_r^2\rangle-\langle G_r\rangle^2\). Expanding \(G_r^2\) gives

\begin{align}
\langle G_r^2\rangle
=
-\frac14\Big[
e^{-2i\vartheta}\langle a^{4}\rangle
+e^{2i\vartheta}\langle a^{\dagger 4}\rangle
-\big\langle a^{2}a^{\dagger 2}+a^{\dagger 2}a^{2}\big\rangle
\Big],
\label{eq:app_Gr2_exp}
\end{align}

while

\begin{align}
\langle G_r\rangle
&=\frac{i}{2}\Big(e^{-i\vartheta}\langle a^2\rangle-e^{i\vartheta}\langle a^{\dagger 2}\rangle\Big),\nn\\
\langle G_r\rangle^2
&=-\frac14\Big[
e^{-2i\vartheta}\langle a^{2}\rangle^{2}
+e^{2i\vartheta}\langle a^{\dagger 2}\rangle^{2}
-2\langle a^{2}\rangle\langle a^{\dagger 2}\rangle
\Big].
\label{eq:app_Gr_mean_sq}
\end{align}

Combining Eqs.~\eqref{eq:app_Gr2_exp} and \eqref{eq:app_Gr_mean_sq} yields

\begin{align}
\mathrm{Var}(G_r)
=
-\frac14\Bigg[
&e^{-2i\vartheta}\,\langle a^{4}\rangle
+e^{2i\vartheta}\,\langle a^{\dagger 4}\rangle
-\Big\langle a^{2}a^{\dagger 2}+a^{\dagger 2}a^{2}\Big\rangle\nn\\
&-e^{-2i\vartheta}\,\langle a^{2}\rangle^{2}
-e^{2i\vartheta}\,\langle a^{\dagger 2}\rangle^{2}
+2\langle a^{2}\rangle\langle a^{\dagger 2}\rangle
\Bigg].
\label{eq:app_Var_Gr_preMN}
\end{align}

To express \(\langle a^{2}a^{\dagger 2}\rangle\) in terms of \(N_1\) and \(N_2\), use

\begin{align}
\langle a^{2}a^{\dagger 2}\rangle
&=\big\langle [a^{2},a^{\dagger 2}]\big\rangle+\langle a^{\dagger 2}a^{2}\rangle,
\label{eq:app_a2ad2_split}
\\
[a^{2},a^{\dagger 2}]
&=a[a,a^{\dagger 2}]+[a,a^{\dagger 2}]a
=2(aa^\dagger+a^\dagger a)
=4a^\dagger a+2,
\label{eq:app_comm_a2ad2}
\end{align}

so that

\begin{align}
\langle a^{2}a^{\dagger 2}\rangle
=N_2+4N_1+2.
\label{eq:app_a2ad2_final}
\end{align}

Using \(\langle a^{\dagger 4}\rangle=M_4^\ast\) and \(\langle a^{\dagger 2}\rangle=M_2^\ast\), Eq.~\eqref{eq:app_Var_Gr_preMN} reduces to

\begin{align}
\mathrm{Var}(G_r)
&=-\frac14\Big(e^{-2i\vartheta}(M_4-M_2^{2})+\mathrm{c.c.}\Big)
+\frac14\Big(2N_2+4N_1+2-2|M_2|^{2}\Big)\nn\\
&=-\frac12\,\Re\Big(e^{-2i\vartheta}(M_4-M_2^{2})\Big)
+\frac12\Big(N_2+2N_1+1-|M_2|^{2}\Big).
\label{eq:app_Var_Gr_final}
\end{align}

\subsection{\texorpdfstring{Variance of $G_{\vartheta}$}{}}

For

\begin{align}
G_{\vartheta}=\frac12\Big(e^{-i\vartheta}a^{2}+e^{i\vartheta}a^{\dagger 2}\Big),
\label{eq:app_Gtheta_def}
\end{align}

a parallel expansion gives

\begin{align}
\mathrm{Var}(G_{\vartheta})
=\frac14\Bigg[
&e^{-2i\vartheta}\,\langle a^{4}\rangle
+e^{2i\vartheta}\,\langle a^{\dagger 4}\rangle
+\Big\langle a^{2}a^{\dagger 2}+a^{\dagger 2}a^{2}\Big\rangle\nn\\
&-e^{-2i\vartheta}\,\langle a^{2}\rangle^{2}
-e^{2i\vartheta}\,\langle a^{\dagger 2}\rangle^{2}
-2\langle a^{2}\rangle\langle a^{\dagger 2}\rangle
\Bigg].
\label{eq:app_Var_Gtheta_preMN}
\end{align}

Using Eqs.~\eqref{eq:app_a2ad2_final} and \eqref{eq:app_MN_defs} then yields

\begin{align}
\mathrm{Var}(G_{\vartheta})
&=\frac14\Big(e^{-2i\vartheta}(M_4-M_2^{2})+\mathrm{c.c.}\Big)
+\frac14\Big(2N_2+4N_1+2-2|M_2|^{2}\Big)\nn\\
&=\frac12\,\Re\Big(e^{-2i\vartheta}(M_4-M_2^{2})\Big)
+\frac12\Big(N_2+2N_1+1-|M_2|^{2}\Big).
\label{eq:app_Var_Gtheta_final}
\end{align}

Multiplying Eqs.~\eqref{eq:app_Var_Gr_final} and \eqref{eq:app_Var_Gtheta_final} by \(4\) reproduces the pure-state QFI expressions in Eqs.~\eqref{eq:QFI_quadratic_strength_moments} and \eqref{eq:QFI_quadratic_angle_moments}, respectively.

\subsection{\texorpdfstring{Janus moment assemblies}{}}
\label{app:janus_moment_assemblies_fixed_squeezing}

For the normalized Janus probe \(\ket{\psi}=\chi\ket{\xi}+\eta\ket{\zeta}\), the anomalous moments needed in Eqs.~\eqref{eq:QFI_quadratic_angle_moments}--\eqref{eq:QFI_quadratic_strength_moments} are assembled as

\begin{align}
M_2
&\equiv \bra{\psi}a^2\ket{\psi}
=
|\chi|^2\,\bra{\xi}a^2\ket{\xi}
+
|\eta|^2\,\bra{\zeta}a^2\ket{\zeta}\nn\\
&\quad
+\chi^\ast\eta\,\bra{\xi}a^2\ket{\zeta}
+\eta^\ast\chi\,\bra{\zeta}a^2\ket{\xi},
\label{eq:M2_Janus_assembly}\\
M_4
&\equiv \bra{\psi}a^4\ket{\psi}
=
|\chi|^2\,\bra{\xi}a^4\ket{\xi}
+
|\eta|^2\,\bra{\zeta}a^4\ket{\zeta}\nn\\
&\quad
+\chi^\ast\eta\,\bra{\xi}a^4\ket{\zeta}
+\eta^\ast\chi\,\bra{\zeta}a^4\ket{\xi},
\label{eq:M4_Janus_assembly}
\end{align}

where the diagonal squeezed-vacuum moments may be written in the \(\alpha,\beta\) notation of Sec.~\ref{sec:janus_quadrature_variances} as
\[
\bra{\xi}a^2\ket{\xi}=-\,\frac{\alpha}{1-|\alpha|^2},
\qquad
\bra{\xi}a^4\ket{\xi}=3\,\frac{\alpha^{2}}{(1-|\alpha|^2)^2},
\]
and similarly with \(\alpha\to\beta\) for \(\ket{\zeta}\).


\section{\texorpdfstring{Equal-strength reduction for $r=s$ and $\Delta\neq 0$}{}}
\label{app:equal_strength_min_DQ}

This appendix records the equal-strength algebra used to obtain Eq.~\eqref{eq:alpha_beta_gamma_equal_strength_main}. Throughout, \(x=\tanh^2 r\) and \(\Delta=\theta-\phi\), with the equal-strength constraint \(r=s\) imposed from the outset.

From Eq.~\eqref{eq:overlap_S_min}, the overlap reduces to

\begin{align}
\mathcal{S}
=(1-x)^{1/2}\,(1-xe^{i\Delta})^{-1/2},
\qquad
|\mathcal{S}|^2=\frac{1-x}{|1-xe^{i\Delta}|},
\label{eq:S_equal_strength_app}
\end{align}

where

\begin{align}
|1-xe^{i\Delta}|=\sqrt{1+x^2-2x\cos\Delta}.
\label{eq:Den_abs_app}
\end{align}

The diagonal elements follow from the single-squeezer moments \(\langle a^\dagger a\rangle=x/(1-x)\) and \(\langle a^2\rangle=-(\sqrt{x}\,e^{i\theta})/(1-x)\), which give

\begin{align}
A_Q
=\frac{1+x-2\sqrt{x}\cos\theta}{2(1-x)},
\qquad
B_Q
=\frac{1+x-2\sqrt{x}\cos\phi}{2(1-x)}.
\label{eq:AB_equal_strength_app}
\end{align}

The off-diagonal element in Eq.~\eqref{eq:CQ_compact} becomes

\begin{align}
C_Q
=(1-x)^{1/2}\,
\frac{\mathcal{N}_Q}{2(1-xe^{i\Delta})^{3/2}},
\qquad
\mathcal{N}_Q\equiv 1+xe^{i\Delta}-\sqrt{x}\,e^{i\theta}-\sqrt{x}\,e^{-i\phi}.
\label{eq:CQ_equal_strength_app}
\end{align}

The combination \(\mathcal{S}^\ast C_Q\) entering \(\alpha\) is simplified by writing \(\mathcal{S}^\ast=(1-x)^{1/2}(1-xe^{-i\Delta})^{-1/2}\), so that

\begin{align}
\mathcal{S}^\ast C_Q
&=\frac{(1-x)\,\mathcal{N}_Q}{2(1-xe^{i\Delta})^{3/2}(1-xe^{-i\Delta})^{1/2}}\nn\\
&=\frac{(1-x)\,\mathcal{N}_Q}{2(1-xe^{i\Delta})\,|1-xe^{i\Delta}|}.
\label{eq:SstarCQ_app_start}
\end{align}

Using the identity

\begin{align}
\frac{1}{1-xe^{i\Delta}}=\frac{1-xe^{-i\Delta}}{|1-xe^{i\Delta}|^2},
\label{eq:inv_den_identity_app}
\end{align}

one obtains

\begin{align}
\Re\big(\mathcal{S}^\ast C_Q\big)
=\frac{(1-x)}{2\,|1-xe^{i\Delta}|^{3}}\,
\Re\Big(\mathcal{N}_Q\,(1-xe^{-i\Delta})\Big).
\label{eq:ReSstarCQ_app_mid}
\end{align}

Imposing \(\Delta=\theta-\phi\) and expanding the real part yields

\begin{align}
\Re\Big(\mathcal{N}_Q\,(1-xe^{-i\Delta})\Big)
=(1-x)\Big[(1+x)-\sqrt{x}\,(\cos\theta+\cos\phi)\Big],
\label{eq:Re_collapse_app}
\end{align}

and therefore

\begin{align}
\Re\big(\mathcal{S}^\ast C_Q\big)
=
\frac{(1-x)^2}{2\,|1-xe^{i\Delta}|^{3}}
\Big[(1+x)-\sqrt{x}\,(\cos\theta+\cos\phi)\Big].
\label{eq:ReSstarCQ_equal_strength_app}
\end{align}

The remaining input for \(\beta\) is \(|C_Q|^2\). From Eq.~\eqref{eq:CQ_equal_strength_app},

\begin{align}
|C_Q|^2
=\frac{(1-x)\,|\mathcal{N}_Q|^2}{4\,|1-xe^{i\Delta}|^{3}}.
\label{eq:CQabs_equal_strength_app}
\end{align}

The modulus \(|\mathcal{N}_Q|^2\) is obtained by multiplying \(\mathcal{N}_Q\) by its complex conjugate and again using \(\Delta=\theta-\phi\). The result can be written in terms of real trigonometric functions of \(\theta\) and \(\phi\) as

\begin{align}
|\mathcal{N}_Q|^2
=(1+x)^2
+4x\cos\theta\cos\phi
-2\sqrt{x}\,(1+x)\,(\cos\theta+\cos\phi).
\label{eq:NQabs_equal_strength_app}
\end{align}

Collecting Eqs.~\eqref{eq:S_equal_strength_app}, \eqref{eq:AB_equal_strength_app}, \eqref{eq:ReSstarCQ_equal_strength_app}, and \eqref{eq:CQabs_equal_strength_app} into the definitions \(\gamma\equiv 1-|\mathcal S|^2\), \(\alpha\equiv(A_Q+B_Q)-2\Re(\mathcal S^\ast C_Q)\), and \(\beta\equiv A_QB_Q-|C_Q|^2\), yields Eq.~\eqref{eq:alpha_beta_gamma_equal_strength_main}. Substituting these expressions into Eq.~\eqref{eq:lambda_Q_pm} then produces the analytic minimum in Eq.~\eqref{eq:DQmin_equal_strength_final} for \(\Delta\neq 0\).

\twocolumngrid

\bibliographystyle{apsrev4-2}
\bibliography{SqueezingRef}
\end{document}